\documentclass[review,12pt]{elsarticle}

\usepackage{hyperref}
\usepackage{changepage}
\journal{Computer Methods in Applied Mechanics and Engineering}
  
\usepackage{amsmath} 
\usepackage{amssymb}
\usepackage{upgreek}
\usepackage{bm} 
\usepackage{mathtools, nccmath} 
\biboptions{sort&compress}
\usepackage{makecell}  
\usepackage{float} 
\usepackage{tablefootnote}
\usepackage{subcaption} 
\usepackage{algorithm} 
\usepackage{algpseudocode}
\usepackage[margin=2cm]{geometry}
\usepackage[capitalise, nameinlink]{cleveref}

\usepackage{soul, xcolor}
\sethlcolor{yellow}

\usepackage{placeins}

\begin{document}

\begin{frontmatter}
\title{A phase field-based framework for electro-chemo-mechanical fracture: crack-contained electrolytes, chemical reactions and stabilisation}

\author{Tim Hageman}
\author{Emilio Martínez-Pañeda \corref{mycorrespondingauthor}}
\cortext[mycorrespondingauthor]{Corresponding author}
\ead{e.martinez-paneda@imperial.ac.uk}

\address{Department of Civil and Environmental Engineering, Imperial College London, London SW7 2AZ, UK}

\begin{abstract}
We present a new theoretical and computational framework for modelling electro-chemo-mechanical fracture. The model combines a phase field description of fracture with a fully coupled characterisation of electrolyte behaviour, surface chemical reactions and stress-assisted diffusion. Importantly, a new physics-based formulation is presented to describe electrolyte-containing phase field cracks, appropriately capturing the sensitivity of electrochemical transport and reaction kinetics to the crack opening height. Unlike other existing methods, this approach is shown to accurately capture the results obtained with discrete fracture simulations. The potential of the electro-chemo-mechanical model presented is demonstrated by particularising it to the analysis of hydrogen embrittlement in metallic samples exposed to aqueous electrolytes. The finite element implementation takes as nodal degrees-of-freedom the electrolyte potential, the concentrations of relevant ionic species, the surface coverage, the concentration of diluted species, the displacement field and the phase field order parameter. Particular attention is devoted to improve stability and efficiency, resulting in the development of strategies for avoiding ill-constrained degrees of freedom and lumped integration schemes that eliminate numerical oscillations. The numerical experiments conducted showcase the ability of the model to deliver assumptions-free predictions for systems involving both free-flowing and crack-contained electrolytes. The results obtained highlight the role of electrolyte behaviour in driving the cracking process, evidencing the limitations of existing models.\\
\end{abstract}


\end{frontmatter}

\section{Introduction}
\label{sec:intro}
Many problems of technological significance are driven by the coupling between electrochemistry and mechanics. In stress corrosion cracking, cracks nucleate and grow through a combination of mechanical loads and corrosion reactions \cite{Sieradzki1987,Turnbull1987,Turnbull2010,Winzer2005,Martinez-Paneda2021}. In the context of Li-Ion batteries, the development of cracks due to chemical strains results in degradation of battery performance and capacity \cite{Tapia-Ruiz2020,Zhao2022,Boyce2022,Ai2022}. In metals exposed to hydrogen-containing environments, one observes a remarkable reduction in ductility and fracture resistance as a result of hydrogen ingress and the associated embrittlement \cite{Gangloff2012}. These sets of problem are characterised by their strongly coupled nature. Take for example the case of metal embrittlement due to hydrogen uptake from aqueous electrolytes. The failure load is governed by the local magnitude of the mechanical fields and of the concentration of hydrogen dissolved in the metal, which are themselves coupled (e.g., hydrogen accumulates in regions of high hydrostatic stress) and dependent on the geometry of the crack. Moreover, hydrogen ingress into the metal is governed by the near-surface stress state, the electrochemical reaction rates at the electrolyte-metal interface, and the electrochemical behaviour of the electrolyte, with all of these items being dependent on the defect geometry while at the same time influencing the defect morphology evolution. For example, the defect dimensions (e.g., crack opening height) will have a major influence on the local chemistry and electrolyte behaviour, which in turn will affect hydrogen uptake. Thus, predicting electro-chemo-mechanical fracture phenomena requires developing models capable of resolving all the coupled physical processes taking place.\\

In this work, we present a new phase field-based model for electro-chemo-mechanical fracture that incorporates all the relevant physical stages. Namely, our theoretical and computational framework handles: (i) the electrochemical behaviour of electrolytes, predicting electrolyte potential distribution and the transport of ionic species due to diffusion and migration; (ii) the chemical reactions occurring at the electrolyte-electrode interface, with the associated kinetic effects and their dependence on electrolyte and surface conditions; (iii) the ingress of diluted species into the material and its diffusion within the solid; (iv) the deformation of the solid, and its coupling with the bulk transport of diluted species; (v) the nucleation and growth of cracks, which can be facilitated by the presence of dilute species; and (vi) a suitable treatment of electrolytes within cracks and other occluded environments. Importantly, computational procedures that can significantly facilitate numerical convergence and stability are also presented . The framework and associated computational schemes are universal but, for demonstration purposes, constitutive choices are made that particularise the model to the simulation of hydrogen embrittlement in metals exposed to aqueous electrolytes. Chemo-mechanical models exist that can predict the transport of dissolved hydrogen within a metal, resolving the interplay between mechanical fields and hydrogen diffusion \cite{Sofronis1989,DiLeo2013,Diaz2016,IJHE2016}. However, these models adopt as boundary condition a constant hydrogen concentration (or chemical potential) at the metal surface, an approach that requires making strong assumptions about the electrolyte conditions, and that has been shown to deliver highly inaccurate predictions \cite{Hageman2022}. More comprehensive, electro-chemo-mechanical models have been recently proposed that attempt at resolving the kinetics of the hydrogen evolution reaction and accurately predict hydrogen ingress by computationally resolving the electrolyte behaviour \cite{Hageman2022,CS2020b}. However, these models only deal with stationary defects. Several methodologies exist to handle propagating cracks and these have been adopted in the hydrogen embrittlement community. Cohesive zone modelling schemes \cite{Serebrinsky2004,Yu2016a,EFM2017}
and phase field fracture models \cite{Martinez-Paneda2018,Wu2020b,Cui2022} have been especially popular. The latter are particularly promising due to their additional modelling capabilities; by indicating the presence of fractured surfaces through an indicator function, the crack path is represented as an additional field, greatly increasing the flexibility and simplicity of the computations \cite{Bourdin2000,Borden2012,Luo2023}. As a result, this approach has gained notable popularity since its development, and has been applied to a large range of materials and damage phenomena such as ductile fracture \cite{Ambati2015a,Miehe2016c,Borden2016}, metallic fatigue \cite{Carrara2020,CMAME2022,Golahmar2023,Alessi2023}, functionally graded materials \cite{CPB2019,Kumar2021}, composites \cite{Reinoso2017,CST2021,Mitrou2023}, shape memory alloys \cite{CMAME2021,Hasan2022}, and iceberg calving \cite{Sun2021,Clayton2022}. While phase field fracture modelling has been widely embraced in the hydrogen embrittlement community (see, e.g. \cite{CS2020,Duda2018,Anand2019,Huang2020,JMPS2020,Mandal2021} and Refs. therein), 
all studies to date require \textit{a priori} knowledge of the hydrogen surface concentration for a given environment. The development of a fully coupled electro-chemo-mechanical framework would eliminate assumptions and deliver predictions purely as a function of the environment, the material and the loading conditions. However, this requires tackling multiple computational challenges.\\

When developing a fully predictive framework for electro-chemo-mechanical fracture, one aspect that requires careful consideration is the treatment of the aqueous electrolyte solution inside of cracked domains. For example, electrolytes acidify in occluded geometries such as pits and cracks, where pH values can change by 80\% depending on the defect geometry, which significantly enhances hydrogen uptake \citep{Kehler2008,Carneiro-Neto2016}. The need to accurately estimate crack openings poses a challenge for smeared modelling approaches such as phase field fracture as the crack is not explicitly represented. Here, one can take inspiration from the work conducted on the area of hydraulic fracture. While the highly conductive fractures assumption is occasionally used when simulating pressurised cracks \cite{Wu2016,Schuler2020}, a more common strategy is to base the diffusivity of the fluid on the opening of the cracks, prescribing fluid fluxes based on simplified relations \cite{Miehe2016b,Lee2016c,Santillan2018,Chukwudozie2019}. These formulations reconstruct the crack opening height to impose a physically realistic fluid flow profile, and thereby increase the accuracy of the overall simulations. While not directly applicable to electrochemical transport within cracks, the manner in which the coupling between the displacement of the solid material and the state of the fluid within cracks is introduced could act as a basis to develop a rigorous scheme for electro-chemo-mechanical simulations. Here, we present a physics-based approach that enables connecting the crack height with the electrolyte behaviour. Other computational aspects, such as the use of a lumped integration scheme for improving stabilisation and robustness, are also discussed.\\

The remainder of this paper is arranged as follows. First, in Section \ref{sec:gov_eq}, we present our electro-chemo-mechanical framework encompassing electrolyte behaviour, surface chemical reactions, hydrogen uptake and diffusion in the metal, mechanical deformation, and a phase field description of fracture with a suitable electrolyte-crack treatment. There, we also introduce our new physics-based approach for describing electrolytes contained within phase field cracks. Then, in Section \ref{sec:implementation}, we describe the numerical implementation of our theory, emphasising the handling of the couplings, convergence criteria, the strategies adopted for the prevention of ill-constrained degrees-of-freedom, and the lumped integration scheme adopted to improve numerical stabilisation. The results obtained are given in Section \ref{sec:results}. First, we examine the abilities of our new physics-based formulation for handling electrolytes within phase field cracks, comparing it with existing approaches \cite{Wu2016} and with discrete fracture simulations. Then, fully coupled electro-chemo-mechanical predictions are obtained for boundary value problems of particular interest, showcasing the ability of the model to capture fracture phenomena involving free-flowing and crack-contained electrolytes. Finally, concluding remarks end the paper in Section \ref{sec:conclusion}. 

\section{A theory for electro-chemo-mechanical fracture}
\label{sec:gov_eq}

\begin{figure}
    \centering
    \includegraphics[width=18cm]{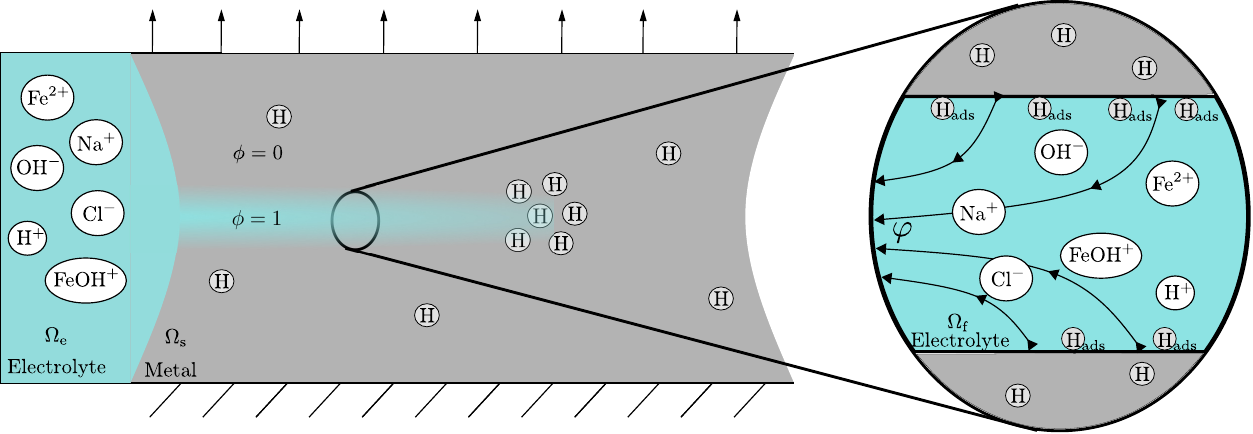}
    \caption{Overview of the coupled electro-chemo-mechanical system, consisting of a metal undergoing deformation, hydrogen absorption and cracking, and an electrolyte containing several ionic species with their movement allowing for electric currents.}
    \label{fig:generic_domain_overview}
\end{figure}

A domain $\Omega$ is considered, consisting of a metal domain $\Omega_\mathrm{s}$, and an electrolyte domain $\Omega_\mathrm{e}$, as shown in \cref{fig:generic_domain_overview}. Cracks are present in the metal domain, with the electrolyte contained within these cracks included as an \textit{ad-hoc} formulation that builds upon the definition of a domain $\Omega_f$ for the crack-electrolyte region. The displacement field in the solid is denoted by $\mathbf{u}$. Within the metal, hydrogen is dissolved at the interstitial lattice sites, with the interstitial lattice hydrogen concentration given by $C_\mathrm{L}$. The presence of fractures is described using a phase field order parameter $\phi$, with $\phi=0$ denoting intact material points and $\phi=1$ a locally fully fractured state. The state of the electrolyte in the $\Omega_\mathrm{f}$ and $\Omega_\mathrm{e}$ domains is described by the variables $\varphi$ and $C_\pi$, which respectively denote the electric potential of the electrolyte and the concentrations of ionic species $\pi=\mathrm{H}^+,\mathrm{OH}^-,\mathrm{Fe}^{2+},\mathrm{FeOH}^+,\mathrm{Na}^+, \mathrm{Cl}^-$. These species are chosen as representative of a conductive electrolyte (e.g., sea water). Finally, the hydrogen coverage of the metal-electrolyte interface is given by $\theta_{\mathrm{ads}}$. Thus, the behaviour of the electro-chemo-mechanical system is described by means of 12 field quantities, which are coupled together through physical phenomena; their governing equations and associated couplings are provided below. \\

We proceed to describe the governing equations describing behaviour in the solid domain (Section \ref{Sec:SolidDomain}) and the electrolyte domain (Section \ref{Sec:Electrolyte}). In addition, the treatment of the electrolyte contained within cracks is extensively discussed in Section \ref{sec:ElectrolyteCrack}, presenting our approach to consistently handle this important aspect of the model.

\subsection{Solid domain}
\label{Sec:SolidDomain}

The solid domain consists of a metal deforming under the small strain assumption, such that the strain tensor $\bm{\varepsilon}$ is given by,
\begin{equation}
  \bm{\varepsilon} =  \frac{1}{2} \left(\nabla^T \mathbf{u} + \nabla \mathbf{u} \right)
\end{equation}

The solid can contain or develop cracks and, accordingly, the total potential energy of the solid includes stored and fracture energy contributions, such that
\begin{equation}
    \Pi = \int_{\Omega_\mathrm{s}} \psi \;\mathrm{d}{\Omega_\mathrm{s}} + \int_{\Gamma_\mathrm{d}} G_\mathrm{c} \;\mathrm{d}\Gamma_\mathrm{d} \label{eq:energy_solid}
\end{equation}
\noindent where $\psi$ is the stored (elastic) strain energy density and $G_\mathrm{c}$ denotes the critical energy release rate or material toughness, which is dependent on the lattice hydrogen concentration; $G_\mathrm{c}(C_\mathrm{L})$. To describe the evolution of cracks, we adopt a smeared representation based on phase field fracture formulations \cite{Bourdin2000,Kristensen2021}. Accordingly, a damage function $d(\phi)$ is defined to capture the degradation of the material and the fracture energy is regularised using a so-called crack density function, which is a function of the phase field and its gradient, $\gamma (\phi,  \nabla \phi) $. Thus, a quantity $\square$ is distributed over a region within $\Omega_\mathrm{s}$, such that
\begin{equation}
    \int_{\Gamma_\mathrm{d}} \square \; \mathrm{d}\Gamma_\mathrm{d} = \int_{\Omega_\mathrm{s}} \gamma \left( \phi,  \bm{\nabla}  \phi \right) \square \; \mathrm{d}{\Omega_\mathrm{s}} \qquad \mathrm{where} \qquad \gamma \left( \phi,  \nabla \phi \right) = \frac{1}{2\ell}\phi^2+\frac{\ell}{2}\left|\bm{\nabla} \phi \right|^2 \label{eq:distributor_phasefield}
\end{equation}
with the length scale $\ell$ controlling the width of this region. Using this distribution function, a regularised form of \cref{eq:energy_solid} can be written as:
\begin{equation}
    \Pi = \int_{\Omega_\mathrm{s}} d(\phi)\psi_{\mathrm{0}} +\gamma \left( \phi,  \nabla \phi \right) G_\mathrm{c}(C_\mathrm{L}) \;\mathrm{d}{\Omega_\mathrm{s}}
\end{equation}
which is then used to obtain the strong forms of momentum balance and fracture evolution of the metal as:
\begin{equation}
    0 = \bm{\nabla} \cdot \frac{\partial \Pi}{\partial \bm{\varepsilon}} = \bm{\nabla}  \cdot \left(d(\phi) \mathcal{D}\bm{:}\bm{\varepsilon}\right) 
    \label{eq:momentumbalance}
\end{equation}
\begin{equation}
    0=\bm{\nabla}\cdot \frac{\partial \Pi}{\partial \bm{\nabla} \phi}+\frac{\partial \Pi}{\partial \phi} = G_\mathrm{c}(C_\mathrm{L}) \left( \frac{1}{\ell}\phi - \ell \bm{\nabla}^2\phi \right) + \frac{\partial d(\phi)}{\partial \phi}\psi_\mathrm{0}
    \label{eq:phasefieldEvolution}
\end{equation}
which assumes the non-fractured metal to behave as a linear-elastic solid with stiffness tensor $\mathcal{D}$. To enforce the irreversibility of the phase field parameter, a history variable $\mathcal H$ is introduced \cite{Miehe2010b}, transforming \cref{eq:phasefieldEvolution} into:
\begin{equation}
   \frac{\phi}{\ell} - \ell \bm{\nabla}^2\phi = - \frac{\partial d(\phi)}{\partial \phi} \mathcal{H} \qquad \mathrm{where} \qquad \mathcal{H} = \frac{\psi_\mathrm{0}}{G_\mathrm{c}(C_\mathrm{L})}, \; \, \dot{\mathcal{H}}>=0
    \label{eq:phasefieldEvolution2}
\end{equation}
Since the fracture energy is a function of the lattice concentration, it is included inside the definition of the history variable to prevent the phase field variable from decreasing when the lattice hydrogen concentration decreases. Throughout this work, a quadratic degradation function will be used, such that
\begin{equation}
    d(\phi) = k_\mathrm{0} + (1-k_\mathrm{0}) (1-\phi)^2 \label{eq:degradeFunction}
\end{equation}
where $k_\mathrm{0}=10^{-10}$ is a \emph{residual stiffness}, introduced to prevent the mechanical sub-problem from becoming ill-posed. Following Martínez-Pañeda et al. \cite{Martinez-Paneda2018}, the degradation of the material toughness with increasing hydrogen concentration is defined as,
\begin{equation}
    G_\mathrm{c}(C_\mathrm{L}) = G_{\mathrm{c0}}\left(1-\chi \frac{C_\mathrm{L}/N_\mathrm{L}}{C_\mathrm{L}/N_\mathrm{L}+\exp(-\Delta g_\mathrm{b}/RT)}\right) \label{eq:Gc}
\end{equation}
\noindent where $G_{\mathrm{c0}}$ is the material toughness in the absence of hydrogen, $\Delta g_\mathrm{b}$ is the binding energy of the critical interfaces, $\chi$ is the hydrogen degradation factor, $N_\mathrm{L}$ is the concentration of interstitial lattice sites, $R$ is the gas constant, and $T$ is the temperature.

It remains to describe the stress-assisted diffusion of hydrogen within the bulk metal. To this end, a dissolved hydrogen chemical potential is defined as,
\begin{equation}
    \mu = \mu_\mathrm{0} + RT \ln\left(\frac{\theta_\mathrm{L}}{1-\theta_\mathrm{L}}\right)-\overline{V}_\mathrm{H}\sigma_\mathrm{H} \label{eq:chemPot}
\end{equation}
where $\mu_0$ is the reference chemical potential, $\theta_\mathrm{L} = C_\mathrm{L}/N_\mathrm{L}$ is the occupancy of interstitial lattice sites, $\overline{V}_\mathrm{H}$ is the partial molar volume of hydrogen, and $\sigma_\mathrm{H}$ is the hydrostatic stress, which is defined as a function of the Cauchy stress tensor as $\sigma_\mathrm{H}=\mathrm{tr}(d(\phi)\bm{\sigma}_\mathrm{0})/3$. Dissolved hydrogen atoms can diffuse freely through the crystal lattice or be sequestered at microstructural trap sites such as carbides, grain boundaries or dislocations \cite{Barrera2016,AM2020}. Accordingly, a concentration of hydrogen in trap sites is defined as $C_T$, such that the total concentration equals $C=C_L+C_T$. Considering that a metal can contain multiple trap types, the mass balance is given by, 
\begin{equation}
    0 = \dot{C}_\mathrm{L} + \sum_\mathrm{traps}\dot{C}_\mathrm{T} + \bm{\nabla} \cdot \mathbf{j}_\mathrm{L} \label{eq:massBalanceHyd}
\end{equation}
where a diffusive flux $\mathbf{j}_\mathrm{L}=-D_\mathrm{L} C_\mathrm{L}/RT \; \bm{\nabla} \mu$ is considered. Assuming equilibrium between the hydrogen in trapping sites and in  interstitial sites, these two concentrations are related through \cite{Oriani1970, Diaz2019}:
\begin{equation}
    \frac{C_T}{N_T} = \frac{\frac{C_L}{N_L}\;\mathrm{exp}\left(\Delta g_T/RT\right)}{1+\frac{C_L}{N_L}\;\mathrm{exp}\left(\Delta g_T/RT\right)} \label{eq:trappedHyd}
\end{equation}
with trapping site concentration $N_\mathrm{T}$, and binding energy of the trapping site $\Delta g_\mathrm{T}$.
Substituting the chemical potential, \cref{eq:chemPot}, and the relation between lattice and trapped hydrogen, \cref{eq:trappedHyd}, into the mass balance equation \cref{eq:massBalanceHyd}, results in
\begin{equation}
\begin{split}
    0 &= \left( 1+\frac{\partial \dot{C}_\mathrm{T}}{\partial \dot{C}_\mathrm{L}}\right)\dot{C_\mathrm{L}}-\bm{\nabla}\cdot\frac{D_\mathrm{L}C_\mathrm{L}}{RT}\bm{\nabla}\mu \\ &= \left( 1+\frac{N_\mathrm{T}/N_\mathrm{L} \exp\left(\Delta g_\mathrm{b}/RT\right)}{\left(C_\mathrm{L}/N_\mathrm{L}+\exp\left(\Delta g_\mathrm{b}/RT\right)\right)^2}\right) \dot{C_\mathrm{L}}-\bm{\nabla}\cdot\left(\frac{D_\mathrm{L}}{1-C_\mathrm{L}/N_\mathrm{L}}\bm{\nabla} C_\mathrm{L}\right) + \bm{\nabla} \cdot \left( \frac{D_\mathrm{L} C_\mathrm{L} \overline{V}_\mathrm{H}}{RT}\bm{\nabla}\sigma_\mathrm{H} \right) \label{eq:HydrogenMassBalance}
    \end{split}
\end{equation}
where $D_\mathrm{L}$ is the lattice diffusion coefficient. For simplicity, we consider only one type of trapping site - grain boundaries, with binding energy $\Delta g_\mathrm{T} = \Delta g_\mathrm{b}$, which are also taken to be the critical interface governing the fracture resistance of the material, \cref{eq:Gc}. 

\subsection{Electrolyte domain}
\label{Sec:Electrolyte}

Let us now consider the equations describing the behaviour of electrolytes. The ions within the electrolyte are conserved using the Nernst-Planck mass balance:
\begin{equation}
    0 = \dot{C}_\pi - \bm{\nabla}\cdot\left(D_\pi \bm{\nabla} C_\pi \right) - \frac{z_\pi F}{RT}\bm{\nabla} \cdot \left(D_\pi C_\pi \bm{\nabla} \varphi\right) + R_\pi \label{eq:NernstPlanck}
\end{equation}
using the Faraday constant $F$ and volume reaction rate $R_\pi$. This describes the transport of each of the $\pi$ ions with diffusion coefficient $D_{\pi}$, driven by gradients in the concentration, and for ions with charge $z_{\pi}$ by gradients in electric potential $\varphi$ within the electrolyte. In addition, the conservation of electric current through the electroneutrality condition is used to provide the $\pi+1$ equation needed \citep{Sarkar2011}:
\begin{equation}
    0 = \sum_\pi z_\pi C_\pi \label{eq:electroneutrality}
\end{equation}

For the reactions within the electrolyte, we include the water auto-ionisation reaction:
\begin{equation}
    \mathrm{H}_2\mathrm{O} \xrightleftharpoons[k_{\mathrm{w}}']{k_{\mathrm{w}}} \mathrm{H}^+ + \mathrm{OH}^- \label{react:water}
\end{equation}
with reaction rates:
\begin{equation}
    R_{\mathrm{H}^+,\mathrm{w}}=R_{\mathrm{OH}^-} = k_{\mathrm{w}}C_{\mathrm{H}_2\mathrm{O}} - k_{\mathrm{w}}'C_{\mathrm{H}^+}C_{\mathrm{OH}^-}  = k_{\mathrm{eq}} \left(K_\mathrm{w}-C_{\mathrm{H}^+} C_{\mathrm{OH}^-} \right) \label{eq:water_react}
\end{equation}
and the hydrolysis of the metal ions:
\begin{equation}
    \mathrm{Fe}^{2+} + \mathrm{H}_2\mathrm{O} \xrightleftharpoons[k_{\mathrm{fe}}']{k_{\mathrm{fe}}} \mathrm{FeOH}^+ + \mathrm{H}^+ \label{react:fe_feoh}
\end{equation}
\begin{equation}
    \mathrm{FeOH}^{+} + \mathrm{H}_2\mathrm{O} \xrightharpoonup{k_{\mathrm{feoh}}} \mathrm{Fe}(\mathrm{OH})_2 + \mathrm{H}^+ \label{react:feoh_feoh2}
\end{equation}
with reaction rates:
\begin{align}
    R_{\mathrm{Fe}^{2+}}&=-k_{\mathrm{fe}}C_{\mathrm{Fe}^{2+}}+k_{\mathrm{fe}}'C_{\mathrm{FeOH}^+}C_{\mathrm{H}^+} \\
    R_{\mathrm{FeOH}^+}&=k_{\mathrm{fe}}C_{\mathrm{Fe}^{2+}}-C_{\mathrm{FeOH}^+}(k_{\mathrm{feoh}}+k_{\mathrm{fe}}'C_{\mathrm{H}^+})\\
    R_{\mathrm{H}^+,\mathrm{fe}}&=k_{\mathrm{fe}}C_{\mathrm{Fe}^{2+}}-C_{\mathrm{FeOH}^+}(k_{\mathrm{fe}}'C_{\mathrm{H}^+}-k_{\mathrm{feoh}})
\end{align}
These reactions use forward and backward reaction constants $k$ and $k'$, with the hydrolysis reactions assumed to occur slowly, while the auto-ionisation reaction is assumed to always be in equilibrium. This equilibrium is enforced in \cref{eq:water_react} by using the equilibrium constant $K_\mathrm{w}=k_\mathrm{w}C_{\mathrm{H}_2\mathrm{O}}/k_\mathrm{w}'=10^{-8}\;\mathrm{mol}^2/\mathrm{m}^3$ and by adopting a sufficiently high penalty-like reaction constant $k_{\mathrm{eq}}$.\\

The reactions between the metal surface and the electrolyte are given through the hydrogen evolution reaction (composed of the Volmer, Tafel, Heyrovsky, and absorption reaction steps) and the corrosion reaction \citep{Hageman2022, Liu2014}:
\begin{alignat}{2}
 \text{Volmer (acid):} && \mathrm{H}^+ + \mathrm{M} + \mathrm{e}^- &\xrightleftharpoons[k_{\mathrm{Va}}']{k_{\mathrm{Va}}} \mathrm{MH}_{\mathrm{ads}} \label{react:1} \\
  \text{Heyrovsky (acid):} && \qquad \mathrm{H}^+ + \mathrm{e}^- + \mathrm{MH}_{\mathrm{ads}}&\xrightleftharpoons[k_{\mathrm{Ha}}']{k_{\mathrm{Ha}}} \mathrm{M} + \mathrm{H}_2 \label{react:2} \\
    \text{Volmer (base):} &&  \mathrm{H}_2\mathrm{O} + \mathrm{M} + \mathrm{e}^- &\xrightleftharpoons[k_{\mathrm{Vb}}']{k_{\mathrm{Vb}}} \mathrm{MH}_{\mathrm{ads}} + \mathrm{OH}^- \label{react:5} \\
   \text{Heyrovsky (base):} && \qquad  \mathrm{H}_2\mathrm{O} + \mathrm{e}^- + \mathrm{MH}_{\mathrm{ads}}&\xrightleftharpoons[k_{\mathrm{Hb}}']{k_{\mathrm{Hb}}} \mathrm{M} + \mathrm{H}_2 + \mathrm{OH}^- \label{react:6} \\
    \text{Tafel:} && 2 \mathrm{MH}_{\mathrm{ads}} &\xrightleftharpoons[k_\mathrm{T}']{k_\mathrm{T}} 2\mathrm{M} + \mathrm{H}_2 \label{react:3} \\
   \text{Absorption:} && \mathrm{MH}_{\mathrm{ads}} &\xrightleftharpoons[k_\mathrm{A}']{k_\mathrm{A}} \mathrm{MH}_{\mathrm{abs}}  \label{react:4} \\
   \text{Corrosion:} && \qquad  \mathrm{Fe}^{2+}+2\mathrm{e}^- &\xrightleftharpoons[k_\mathrm{c}']{k_\mathrm{c}} \mathrm{Fe} \label{react:7}
\end{alignat}
with their respective forward and backward reaction rates given by:
\begin{fleqn}[-1.25cm]
\begin{alignat}{4}
\nonumber \hspace{-1cm} && && & \qquad\mathrm{Forward} &&  \qquad\qquad \mathrm{Backward} \\
    \hspace{-1.25cm} \mathrm{Volmer (acid):} && \quad && \nu_{\mathrm{Va}} &= k_{\mathrm{Va}} C_{\mathrm{H}^+}(1-\theta_{\mathrm{ads}})\exp \left( {-\alpha_{\mathrm{Va}} \frac{\eta F}{RT}}\right)\qquad
    && \nu_{\mathrm{Va}}' = k_{\mathrm{Va}}' \theta_{\mathrm{ads}}\exp \left({(1-\alpha_{\mathrm{Va}}) \frac{\eta F}{RT}}\right) \label{eq:react1}\\
    \hspace{-1.25cm} \mathrm{Heyrovsky (acid):} && && \nu_{\mathrm{Ha}} &= k_{\mathrm{Ha}} C_{\mathrm{H}^+}\theta_{\mathrm{ads}}\exp \left({-\alpha_{\mathrm{Ha}} \frac{\eta F}{RT}}\right)\qquad
    && \nu_{\mathrm{Ha}}' = k_{\mathrm{Ha}}' (1-\theta_{\mathrm{ads}}) p_{\mathrm{H}_2} \exp \left({(1-\alpha_{\mathrm{Ha}}) \frac{\eta F}{RT}}\right) \hspace{-1.25cm} \label{eq:react2}\\
    \hspace{-1.25cm} \mathrm{Volmer (base):} && && \nu_{\mathrm{Vb}} &= k_{\mathrm{Vb}} (1-\theta_{\mathrm{ads}})\exp \left({-\alpha_{\mathrm{Vb}} \frac{\eta F}{RT}}\right)\qquad
    && \nu_{\mathrm{Vb}}' = k_{\mathrm{Vb}}' C_{\mathrm{OH}^-} \theta_{\mathrm{ads}}\exp \left({(1-\alpha_{\mathrm{Vb}}) \frac{\eta F}{RT}}\right) \label{eq:react5}\\
    \hspace{-1.25cm} \mathrm{Heyrovsky (base):} && && \nu_{\mathrm{Hb}} &= k_{\mathrm{Hb}} \theta_{\mathrm{ads}}\exp \left({-\alpha_{\mathrm{Hb}} \frac{\eta F}{RT}}\right)\qquad
    && \nu_{\mathrm{Hb}}' = k_{\mathrm{Hb}}' (1-\theta_{\mathrm{ads}}) p_{\mathrm{H}_2} C_{\mathrm{OH}^-} \exp \left({(1-\alpha_{\mathrm{Hb}}) \frac{\eta F}{RT}}\right)  \hspace{-1.25cm} \label{eq:react6}\\
    \hspace{-1.25cm} \mathrm{Tafel:} && && \nu_\mathrm{T} &= k_\mathrm{T}\left|\theta_{\mathrm{ads}}\right|\theta_{\mathrm{ads}}\qquad
    && \nu_\mathrm{T}' = k_\mathrm{T}' (1-\theta_{\mathrm{ads}})p_{\mathrm{H}_2} \label{eq:react3}\\
    \hspace{-1.25cm} \mathrm{Absorption:} && && \nu_\mathrm{A} &= k_\mathrm{A} (N_\mathrm{L} - C_\mathrm{L})\theta_{\mathrm{ads}}\qquad
    && \nu_\mathrm{A}' = k_\mathrm{A}' C_\mathrm{L} (1-\theta_{\mathrm{ads}}) \label{eq:react4}\\
    \hspace{-1.25cm} \mathrm{Corrosion:} && && \nu_{\mathrm{c}} &= k_{\mathrm{c}} C_{\mathrm{Fe}^{2+}}\exp \left({-\alpha_{\mathrm{c}} \frac{\eta F}{RT}}\right) \qquad && \nu_{\mathrm{c}}' = k_{\mathrm{c}}' \exp \left({(1-\alpha_{\mathrm{c}}) \frac{\eta F}{RT}}\right)   \label{eq:react7}
\end{alignat}
\end{fleqn}
These rates use reaction rate constants $k$ and $k'$, charge transfer coefficients $\alpha$, and the electric overpotential $\eta$, which is defined as the difference between the potential jump and the equilibrium potential of the specific reaction, $\eta = E_\mathrm{m}-\varphi-E_{\mathrm{eq},\mathrm{H}}$ (using the imposed metal potential $E_\mathrm{m}$, and either the equilibrium potential at reference conditions for the hydrogen reaction, $E_{\mathrm{eq},\mathrm{H}}$ or the corrosion reaction $E_{\mathrm{eq},\mathrm{Fe}}$). Finally, to conserve the hydrogen between the electrolyte and the metal, the mass balance of the surface adsorbed hydrogen is used:
\begin{equation}
    N_{\mathrm{ads}} \dot{\theta}_{\mathrm{ads}} - (\nu_{\mathrm{Va}}-\nu_{\mathrm{Va}}') + \nu_{\mathrm{Ha}} + 2 \nu_\mathrm{T} + (\nu_\mathrm{A}-\nu_\mathrm{A}') - (\nu_{\mathrm{Vb}}-\nu_{\mathrm{Vb}}') + \nu_{\mathrm{Hb}} = 0
    \label{eq:massbalanceinterface}
\end{equation}
and at the metal-electrolyte interface, the $\mathrm{H}^+$, $\mathrm{OH}^-$, $\mathrm{Fe}^{2+}$ fluxes are prescribed on the electrolyte and the lattice hydrogen flux on the metal:
\begin{align}
    \nu_{\mathrm{H}^+} &= -(\nu_{\mathrm{Va}} - \nu_{\mathrm{Va}}') - \nu_{\mathrm{Ha}}  \\
   \nu_{\mathrm{OH}^-} &= \nu_{\mathrm{Vb}} - \nu_{\mathrm{Vb}}' + \nu_{\mathrm{Hb}}   \\
    \nu_{\mathrm{Fe}^{2+}} &= \nu_{\mathrm{c}}'-\nu_\mathrm{c}\\
    \nu_\mathrm{L} &= \nu_\mathrm{A} - \nu_\mathrm{A}'
\end{align}

The governing equations described in this subsection assume that the electrolyte is a well-defined and separate phase relative to the metal, and thus refer to the electrolyte domain $\Omega_\mathrm{e}$. However, for the crack-contained electrolyte, represented via $\Omega_\mathrm{f}$ in \cref{fig:generic_domain_overview} this is not the case. This will be addressed in the next subsection.

\subsection{Treatment of electrolyte within cracks}
\label{sec:ElectrolyteCrack}

In a smeared approach like the electro-chemo-mechanical phase field framework presented here, the metal and the electrolyte coexist when $\phi>0$. This requires establishing relationships between the metal and the electrolyte as a function of the phase field, so as to capture the influence of cracking on electrolyte transport and reactions. A particularly popular approach in this regard is the \emph{distributed diffusion} model developed by Wu and De Lorenzis \cite{Wu2016}, which captures the enhanced electrolyte transport through cracks by enhancing diffusivity in $\phi>0$ regions. Here, we present a new approach, henceforth referred to as the \emph{physics-based} model, which is able to capture sensitivity to the crack height and naturally establishes a link with the discrete problem without any additional parameters. Both models are described and compared below.\\

Common to both the distributed diffusion and physics-based models is the fact that the electrolyte-specific variables can become active in regions of $\Omega_s$, depending on the evolution of the phase field. These electrolyte-specific variables are thus numerically considered in the entire domain, as is common in phase field approaches, but they only have physical meaning in material points experiencing damage, $\phi>0$, where micro- and macro-cracks that can contain the electrolyte are present. 

\subsubsection{Distributed diffusion model}
\label{sec:ModelWuLorenzis}

The distributed diffusion model by Wu and De Lorenzis \cite{Wu2016} captures the enhanced transport of ions through cracks by defining an effective diffusion coefficient tensor that has two main components;
\begin{equation}
    \bm{D}_{\mathrm{eff},\pi} = \bm{D}_{\pi,1} p (1-\phi)^m + \bm{D}_{\pi,2} \phi^m
\end{equation}
The first term accounts for the characteristics of diffusion in porous materials, through a factor $p$ and $\bm{D}_{\pi,1}=D_{\pi,1}\bm{I}$. In the materials of relevance for this study (metals), $p=0$. The second term accounts for the anisotropy in diffusivity that results from the presence of cracks. This is accomplished through the definition of a parameter $m$, to be fitted to experiments, and sufficiently large values of the diffusion coefficient matrix $\bm{D}_{\pi,2}$. For the capacity term, it is often assumed to be independent of the fracture state when the material is porous \citep{Wu2016}. However, since we are considering a non-porous material, we elect to distribute the capacity term consistently with the diffusion term. This results in the following weak form formulation of the mass balance given in \cref{eq:NernstPlanck}: 
\begin{equation}
    0 = \int_{\Omega_\mathrm{s}} \phi^m \dot{C}_{\pi}\delta C - \bm{\nabla} \cdot \left(\phi^m \bm{D}_{\pi,2} \bm{\nabla} C_\pi \right)\delta C - \frac{z_\pi F}{RT}\bm{\nabla} \cdot \left(\bm{D}_{\pi,2} C_\pi \phi^m \bm{\nabla} \varphi \right)\delta C + R_\pi \phi^m \delta C \;\mathrm{d}{\Omega_\mathrm{s}} + \int_{\Gamma_\mathrm{d}^\pm} \nu_\pi \delta C \;\mathrm{d}\Gamma_\mathrm{d}^\pm
    \label{eq:weak_intermediate}
\end{equation}
Finally, distributing the surface-based reactions through the interface distribution function $\gamma$ (\cref{eq:distributor_phasefield}) transforms the last term of  \cref{eq:weak_intermediate} into a domain-wide integral:
\begin{equation}
    \int_{\Gamma_\mathrm{d}^\pm} \nu_\pi \delta C \;\mathrm{d}\Gamma_\mathrm{d}^\pm = \int_{\Omega_\mathrm{s}} 2\gamma \nu_\pi \delta C \;\mathrm{d}{\Omega_\mathrm{s}}
    \label{eq:weak_distribution_reactions}
\end{equation}
where the factor 2 is introduced to account for the fact that the two fracture surfaces react with the electrolyte. From \cref{eq:weak_intermediate,eq:weak_distribution_reactions}, the strong form for the mass balance distributed over the domain $\Omega_\mathrm{s}$ is extracted as:
\begin{equation}
    0 = \phi^m \dot{C}_{\pi} - \bm{\nabla} \cdot \left(\phi^m \bm{D}_{\pi,2}\bm{\nabla} C_{\pi} \right) - \frac{z_\pi F}{RT}\bm{\nabla} \cdot \left(\bm{D}_{\pi,2}C_\pi \phi^m \bm{\nabla} \varphi \right)+\varphi^m R_\pi + \left(\frac{1}{\ell}\phi^2+\ell\left|\bm{\nabla} \phi \right|^2\right) \nu_\pi \label{eq:WuLorenzis_nernstPlanck}
\end{equation}
One thing to note here is that this equation becomes trivial for the case of a non-fractured domain where $\phi=0$, resulting in the local solution for the concentration becoming undefined. This will be discussed in \cref{sec:fix_unconstrained_dofs}. In a similar manner, the weak form for the electroneutrality condition, \cref{eq:electroneutrality}, is obtained as:
\begin{equation}
    0 = \phi^m \sum_\pi z_\pi C_\pi \label{eq:WuLorenzis_Electroneutrality} 
\end{equation}
Together, \cref{eq:WuLorenzis_nernstPlanck,eq:WuLorenzis_Electroneutrality} describe the behaviour of the electrolyte potential and ion concentrations within the domain.

\subsubsection{Physics-based model}
\label{sec:ModelPhysicsBased}

\begin{figure}
     \centering
     \begin{subfigure}[b]{0.32\textwidth}
         \centering
         \includegraphics{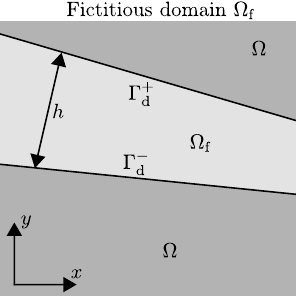}
         \caption{}
         \label{fig:DomainsForDerivation_Fluid}
     \end{subfigure}
     \hfill
     \begin{subfigure}[b]{0.32\textwidth}
         \centering
         \includegraphics{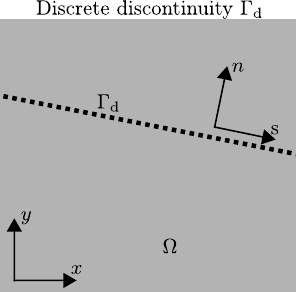}
         \caption{}
         \label{fig:DomainsForDerivation_Discontinuity}
     \end{subfigure}
     \hfill
     \begin{subfigure}[b]{0.32\textwidth}
         \centering
         \includegraphics{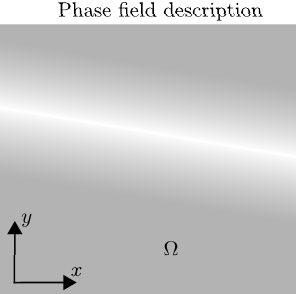}
         \caption{}
         \label{fig:DomainsForDerivation_Phasefield}
     \end{subfigure}
        \caption{Different domains and representations used in the derivation of the physics-based model: (a) fictitious electrolyte domain $\Omega_\mathrm{f}$, (b) discrete discontinuity representation in the domain $\Omega$, and (c) phase field representation in the domain $\Omega$.}
        \label{fig:DomainsForDerivation}
\end{figure}

We have built our physics-based model by considering a fictitious domain $\Omega_\mathrm{f}$, which represents the electrolyte contained within a fracture with opening height $h$, as shown in \cref{fig:DomainsForDerivation_Fluid}. In this fictitious domain, the weak form of the Nernst-Planck mass balance, \cref{eq:NernstPlanck}, is given as:
\begin{equation}
        0 = \int_{\Omega_\mathrm{f}} \dot{C}_{\pi}\delta C - \bm{\nabla} \cdot \left(D_{\pi} \bm{\nabla} C_\pi \right)\delta C - \frac{z_\pi F}{RT}\bm{\nabla} \cdot \left(D_{\pi} C_\pi \bm{\nabla} \varphi \right)\delta C + R_\pi \delta C \;\mathrm{d}\Omega_\mathrm{f} + \int_{\Gamma_\mathrm{d}^\pm} \nu_\pi \delta C \;\mathrm{d}\Gamma_\mathrm{d}^\pm
    \label{eq:weak_fictitious}
\end{equation}
We assume long and thin cracks, as it is commonly the case in corrosive and hydrogen-containing environments, allowing the integrals over domain $\Omega_f$ to be transferred to the discrete discontinuity description of domain $\Omega$ (\cref{fig:DomainsForDerivation_Discontinuity}) through:
\begin{equation}
    \int_{\Omega_\mathrm{f}} \square \;\mathrm{d}\Omega_\mathrm{f} = \int_{\Gamma_\mathrm{d}} h\square \;\mathrm{d}\Gamma_\mathrm{d} \quad, \quad \int_{\Omega_\mathrm{f}} \bm{\nabla} \cdot \square \;\mathrm{d}\Omega_\mathrm{f} = \int_{\Gamma_\mathrm{d}} \frac{\partial}{\partial s} \left(h\square\right) \;\mathrm{d}\Gamma_\mathrm{d} \quad \mathrm{and} \quad \int_{\Gamma_\mathrm{d}^\pm} \square \; \mathrm{d}\Gamma_\mathrm{d}^\pm = \int_{\Gamma_\mathrm{d}} 2\square \;\mathrm{d}\Gamma_\mathrm{d} 
\end{equation}
using the fracture opening height $h$. This allows the weak form for the electrolyte to be defined solely on the discrete discontinuity as:
\begin{equation}
        0 = \int_{\Gamma_\mathrm{d}} h\dot{C}_{\pi}\delta C - \frac{\partial}{\partial s} \left(h D_{\pi} \frac{\partial C_\pi}{\partial s} \right)\delta C - \frac{z_\pi F}{RT}\frac{\partial}{\partial s} \left(h D_{\pi} C_\pi \frac{\partial \varphi}{\partial s} \right)\delta C + h R_\pi \delta C + 2\nu_\pi \delta C \;\mathrm{d}\Gamma_\mathrm{d}
    \label{eq:weak_discontinuity}
\end{equation}
where the terms containing gradients are transformed into unidirectional derivatives along the discontinuity due to the assumption of negligible changes in the direction normal to it. Finally, since the complete weak form is defined on the fracture face, the phase field distribution function from \cref{eq:distributor_phasefield} can be used to distribute the weak form over the complete domain $\Omega$, resulting in:
\begin{equation}
    0 = \int_{\Omega_\mathrm{s}} \gamma h\dot{C}_{\pi}\delta C - \bm{\nabla} \cdot \left(\gamma h \bm{D}_{\pi} \bm{\nabla} C_\pi \right)\delta C - \frac{z_\pi F}{RT}\bm{\nabla} \cdot \left(\gamma  h \bm{D}_{\pi} C_\pi \bm{\nabla} \varphi \right)\delta C + h \gamma R_\pi \delta C + 2\gamma \nu_\pi \delta C \;\mathrm{d}{\Omega_\mathrm{s}}
    \label{eq:weak_Domain} 
\end{equation}
where the assumption of a constant field normal to the fracture is included in the diffusivity matrix $\bm{D}_\pi$. This is achieved by constructing the diffusivity in crack-aligned coordinates (using a rotation matrix $\bm{R}$) as:
\begin{equation}
    \mathbf{R}\bm{D}_\pi\mathbf{R}^T=\begin{bmatrix} D_\pi & 0 \\ 0 & D_\infty/h \end{bmatrix} 
\end{equation}
which assigns the diffusivity of the ionic species $D_\pi$ to the direction tangential to the crack, while notably enhancing diffusion in the normal direction. As this high diffusivity gets multiplied by $\gamma h$, it disappears when the metal is not cracked (where $\gamma h = 0$), while it is constant (and independent of the crack opening height) when the metal is cracked, enforcing negligible concentration gradients normal to the crack. As a result, the concentration within the phase field description will approximate a one-dimensional diffusion along the crack path, consistent with the description of the Nernst-Planck equations for narrow cracks.\\

Based on Eq. (\ref{eq:weak_Domain}), the mass balance for the ion species within the electrolyte is given in its strong form as:
\begin{equation}
     0 = \gamma h\dot{C}_{\pi} - \bm{\nabla} \cdot \left(\gamma h \bm{D}_{\pi} \bm{\nabla} C_\pi \right) - \frac{z_\pi F}{RT}\bm{\nabla} \cdot \left(\gamma  h \bm{D}_{\pi} C_\pi \bm{\nabla} \varphi \right) + h \gamma R_\pi  + 2\gamma \nu_\pi 
    \label{eq:PhysicsBased_nernstPlanck}    
\end{equation}
Accordingly, the electroneutrality condition reads:
\begin{equation}
    0 = \gamma h \sum_\pi z_\pi C_\pi \label{eq:PhysicsBased_Electroneutrality} 
\end{equation}
While the expression for the surface adsorbed hydrogen mass balance, \cref{eq:massbalanceinterface}, is reformulated to
\begin{equation}
    2\gamma\left(N_{\mathrm{ads}} \dot{\theta}_{\mathrm{ads}} - (\nu_{\mathrm{Va}}-\nu_{\mathrm{Va}}') + \nu_{\mathrm{Ha}} + 2 \nu_\mathrm{T} + (\nu_\mathrm{A}-\nu_\mathrm{A}') - (\nu_{\mathrm{Vb}}-\nu_{\mathrm{Vb}}') + \nu_{\mathrm{Hb}}\right) = 0
    \label{eq:PhysicsBased_massbalanceinterface}
\end{equation}
Here, one should note that while the surface $\theta_{\mathrm{ads}}$ is defined in the entire domain, consistent with a phase field description, it only becomes physically meaningful at the electrolyte-metal interface. Thus, when no surfaces are present, the surface coverage is kept at zero, with the method used to enforce this being described in \cref{sec:fix_unconstrained_dofs}.\\

\subsubsection{Estimating the opening heights}
\label{sec:OpeningHeight}
\begin{figure}
    \centering
    \includegraphics{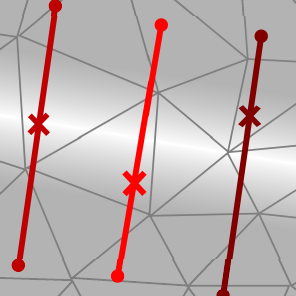}
    \caption{Schematic overview of the lines normal to the crack used to determine the crack opening height for three integration points, which are indicated by red crosses.}
    \label{fig:OpeningHeight}
\end{figure}

The model described in the previous section requires the crack opening height $h$. This opening height is obtained based on the phase field and displacements following the procedure from \citep{Yoshioka2020, Chukwudozie2019}. That is, for every integration point, a surface normal vector is computed as
\begin{equation}
    \bm{n} = \frac{\bm{\nabla}\phi}{\left|\bm{\nabla}\phi\right|}
\end{equation}
which produces a vector normal to the phase field representation of the crack. Using this vector, a line is constructed which crosses the full width of the phase field zone and passes through the integration point currently being considered, as shown in \cref{fig:OpeningHeight}. For the current integration point being considered, the crack opening height is then obtained by integrating along this line as:
\begin{equation}
    h = \int \mathbf{u}\cdot\bm{\nabla}\phi \; \mathrm{d}n \label{eq:heightmapping}
\end{equation}

In order to calculate these integrals, the displacements and phase field gradients are first calculated at all integration points within the domain. This integration point data is then combined to create scattered interpolation functions \citep{ScatteredInterpolant1, ScatteredInterpolant2}, which given the coordinates of an arbitrary point within the domain return the displacements and gradients based on linear interpolation between the closest integration points. Using these interpolation functions, the integral from \cref{eq:heightmapping} is evaluated using numerical integration, obtaining the opening height in the current integration point. This integration is then repeated for \textit{all} integration points within the complete domain, and opening heights are recalculated each time the staggered solution scheme begins solving for the chemical sub-problem (see \cref{alg:solution_scheme}). It should be noted that, as the phase field gradients are discontinuous between neighbouring elements, the use of scattered interpolants to calculate this opening height is not exact, as opposed to directly evaluating the gradients at required locations using the finite element shape functions. However, this method does allow for unstructured meshes and three-dimensional cases to easily be considered, without the need for computationally costly element searches during the integration step. 

\subsubsection{Model comparison}
Comparing the distributed diffusion model, \cref{eq:WuLorenzis_nernstPlanck,eq:WuLorenzis_Electroneutrality}, with the physics-based model, \cref{eq:PhysicsBased_nernstPlanck,eq:PhysicsBased_massbalanceinterface}, shows that both can be cast into the following general form:
\begin{align}
    0 &= \beta_\mathrm{c} \dot{C}_{\pi} - \bm{\nabla} \cdot \left(\bm{R}^T\bm{\beta_\mathrm{d}}\bm{R} D_{\pi} \bm{\nabla} C_\pi \right) - \frac{z_\pi F}{RT}\bm{\nabla} \cdot \left(\bm{R}^T\bm{\beta_\mathrm{d}}\bm{R} D_{\pi} C_\pi \bm{\nabla} \varphi \right) + \beta_\mathrm{c} R_\pi  + \beta_\mathrm{s} \nu_\pi \label{eq:electrolyte_nernstplanck}\\
    0 &= \beta_\mathrm{c} \sum_\pi z_\pi C_\pi \label{eq:electrolyte_neutrality} \\
    0 & =\beta_\mathrm{s} \left(N_{\mathrm{ads}} \dot{\theta}_{\mathrm{ads}} - (\nu_{\mathrm{Va}}-\nu_{\mathrm{Va}}') + \nu_{\mathrm{Ha}} + 2 \nu_\mathrm{T} + (\nu_\mathrm{A}-\nu_\mathrm{A}') - (\nu_{\mathrm{Vb}}-\nu_{\mathrm{Vb}}') + \nu_{\mathrm{Hb}}\right) \label{eq:electrolyte_surface}
\end{align}
using the rotation matrix $\bm{R}$ to align the diffusion within the fracture to its orientation. The capacity, diffusion, and surface distributors are accordingly defined as:
\begin{alignat}{3}
\nonumber & \mathrm{Distributed}\; \mathrm{diffusion} &&  \qquad\qquad && \mathrm{Physics}\mathrm{-}\mathrm{based} \\
&\beta_\mathrm{c} = \phi^m && && \beta_\mathrm{c} = h \left( \frac{1}{2\ell}\phi^2+\frac{\ell}{2}\left|\bm{\nabla} \phi \right|^2 \right) \label{eq:betac} \\
&\bm{\beta}_\mathrm{d} = \begin{bmatrix} \phi^m D_{\pi,2}/D_\pi & 0\\ 0 & 0 \end{bmatrix} && && \bm{\beta}_\mathrm{d} = \left( \frac{1}{2\ell}\phi^2+\frac{\ell}{2}\left|\bm{\nabla} \phi \right|^2 \right) \begin{bmatrix} h & 0\\ 0 & D_{\infty} \end{bmatrix} \\
&\beta_\mathrm{s} = \frac{1}{\ell}\phi^2+\ell\left|\bm{\nabla} \phi \right|^2  && && \beta_\mathrm{s} = \frac{1}{\ell}\phi^2+\ell\left|\bm{\nabla} \phi \right|^2 \label{eq:betas}
\end{alignat}
The main differences between both models can be readily seen upon inspection of Eqs. (\ref{eq:betac})-(\ref{eq:betas}). First, the distributed diffusion model requires calibration of two additional parameters: the exponential factor $m$ and the enhanced diffusion within the fracture $D_{\pi,2}$. These would be expected to have a sensitivity to the crack opening height $h$. In contrast, the description of diffusion in the direction tangential to the crack does not depend on any empirical parameters and naturally incorporates the role of $h$. In this regard, it should be noted that the focus on the physics-based model is not to accurately describe the kinetics of fluid flow within a propagating crack but to accurately capture electrolyte behaviour and its sensitivity to the crack geometry, as crack growth can be highly sensitive to the electrochemical conditions that arise in occluded geometries such as cracks. A second main difference between the two models lies in their description of the transport in the direction normal to the fracture, with the distributed diffusion model assuming that the presence of cracks does not contribute to this transport, while one of the assumptions of the physics-based model is the absence of any gradients normal to the fracture, which is enforced by assigning a large value to $D_\infty$.

\section{Numerical implementation}
\label{sec:implementation}

\begin{algorithm}[t]
\caption{Overview of solution method}\label{alg:solution_scheme}
\begin{algorithmic}[1]
\State Start of time increment
\While{not converged}
\State Solve \cref{eq:phasefieldEvolution2} to update $\bm{\upphi}$
\State Solve \cref{eq:momentumbalance} to update $\mathbf{u}$
\State Update $h$ using \cref{eq:heightmapping}
\While{$\mathbf{C}_\mathrm{L}$, $\mathbf{C}_\pi$, $\bm{\upvarphi}$, $\bm{\uptheta}$ are not converged}
    \State Solve \cref{eq:electrolyte_nernstplanck,eq:electrolyte_neutrality,eq:electrolyte_surface,eq:HydrogenMassBalance} to update $\mathbf{C}_\mathrm{L}$, $\mathbf{C}_\pi$, $\bm{\upvarphi}$, $\bm{\uptheta}$
    \State Calculate energy based residual for $\mathbf{C}_\mathrm{L}$, $\mathbf{C}_\pi$, $\bm{\upvarphi}$, $\bm{\uptheta}$
\EndWhile
    \State Calculate energy based residual for $\mathbf{u}$ and $\bm{\upphi}$
\EndWhile
\State Go to next time increment
\end{algorithmic}
\end{algorithm}

The governing equations, \cref{eq:electrolyte_nernstplanck,eq:electrolyte_neutrality,eq:electrolyte_surface,eq:HydrogenMassBalance,eq:phasefieldEvolution2,eq:momentumbalance}, are solved using the finite element method. The system of equations is solved in an iteratively staggered manner. First, a solution for the phase field is obtained by solving \cref{eq:phasefieldEvolution2}. Next, the displacements are updated by solving \cref{eq:momentumbalance}. Then, a Newton-Raphson solver is used to iteratively solve \cref{eq:electrolyte_nernstplanck,eq:electrolyte_neutrality,eq:electrolyte_surface,eq:HydrogenMassBalance} in a concurrent fashion, so as to update the electrolyte potential, the concentrations of ionic species, the surface coverage, and the hydrogen lattice concentration. Once all the fields have been updated, the convergence of the total system of equations is evaluated, with global iterations taking place until convergence is reached. The solution process is summarised in \cref{alg:solution_scheme}. By using this staggered scheme, we avoid the well-known convergence difficulties that arise when simultaneously solving for the phase field and displacement \citep{Miehe2010b}. Additionally, the non-local mapping for the fracture height, \cref{eq:heightmapping}, only needs to be conducted once per electro-chemical solution step since the displacements and phase field parameter are constant during its solution process. 

\subsection{Spatial and temporal discretisation}
The temporal discretisation of the governing equations is performed using a backward Euler scheme. For the spatial discretisation, quadratic elements are used to discretise the degrees of freedom as:
\begin{equation}
  \begin{alignedat}{3}
    \mathbf{u} &= \sum \bm{N}_\mathrm{u}^{\mathrm{el}} \mathbf{u}^{\mathrm{el}} \qquad && \phi = \sum \mathbf{N}_\phi^{\mathrm{el}} \bm{\upphi}^{\mathrm{el}} \qquad && C_\mathrm{L} = \sum \mathbf{N}_\mathrm{L}^{\mathrm{el}} \mathbf{C}_\mathrm{L}^{\mathrm{el}} \\
    \varphi &= \sum \mathbf{N}_\varphi^{\mathrm{el}} \bm{\upvarphi}^{\mathrm{el}} && \theta_{\mathrm{ads}} = \sum \mathbf{N}_\theta^{\mathrm{el}} \bm{\uptheta}^{\mathrm{el}} && C_\pi = \sum \mathbf{N}_\mathrm{C}^{\mathrm{el}} \mathbf{C}_\pi^{\mathrm{el}}
  \end{alignedat}
\end{equation}
One thing to note about the use of these quadratic elements is the requirement in \cref{eq:HydrogenMassBalance} of second-order derivatives to calculate the hydrostatic stress gradient. These gradients are poorly defined using quadratic $C^0$ inter-element continuous shape functions. While this could be resolved using elements with a higher inter-element continuity, for instance using NURBS \citep{Hughes2005,Borden2011}, T-splines \citep{Bazilevs2010, Scott2011, Hageman2021c}, or Hermitian polynomials \citep{Bogner1965, Ciarlet1972}, no issues were encountered during simulations provided a sufficiently fine mesh was used to discretise the displacement. As phase field methods already impose requirements on the maximum element size, the meshes adopted were sufficiently fine to accurately characterise the hydrostatic stress gradients. 

\subsection{Residuals and stiffness matrices}

We proceed to formulate the residuals and stiffness matrices for each of the governing fields and associated balance equations. 

\subsubsection{Phase field evolution sub-problem}
The first step of the staggered solution scheme is solving the phase field evolution, \cref{eq:phasefieldEvolution2}. This equation is given in discretised weak form as:
\begin{equation}
\begin{split}
    \mathbf{f}_\phi^{t+\Delta t} = &\int_{\Omega_\mathrm{s}} \frac{1}{\ell} \mathbf{N}_\phi^T\mathbf{N}_\phi \bm{\upphi}^{t+\Delta t} + \ell \left(\bm{\nabla}\mathbf{N}_\phi\right)^T\bm{\nabla}\mathbf{N}_\phi \bm{\upphi}^{t+\Delta t} - 2 \left(1-k_\mathrm{0}\right)\mathbf{N}_\phi^T \left(1-\mathbf{N}_\phi \bm{\upphi}^{t+\Delta t}\right) \mathcal{H}^{t+\Delta t} \; \mathrm{d}{\Omega_\mathrm{s}} \\ &- \int_{\Gamma} \ell \bm{N}_\phi^T\bm{\nabla} \mathbf{N}_\phi \bm{\upphi}^{t+\Delta t} \cdot \mathbf{n} \; \mathrm{d}\Gamma
\end{split}
\end{equation}
where the last term, the boundary condition $\nabla \phi \cdot \mathbf{n}$, is set equal to zero hereafter. The discretised history variable is defined as:
\begin{equation}
    \mathcal{H}^{t+\Delta t} = \max \left( \mathcal{H}^t, \;\; \frac{\frac{1}{2} {\mathbf{u}^{t+\Delta t}}^T\bm{B}_\mathrm{u}^T\bm{D}\bm{B}_\mathrm{u}\mathbf{u}^{t+\Delta t}}{G_{\mathrm{c0}} \left( 1-\chi \frac{\mathbf{N}_\mathrm{L} \mathbf{C}_\mathrm{L}^{t+\Delta t}/N_\mathrm{L}}{\mathbf{N}_\mathrm{L} \mathbf{C}_\mathrm{L}^{t+\Delta t}/N_\mathrm{L}+\exp\left(-\Delta g_\mathrm{b}/RT\right)}\right)}\;\;\;\right)
\end{equation}
which uses the displacement to strain mapping matrix $\mathbf{\upvarepsilon}=\bm{B}_\mathrm{u} \mathbf{u}$ and the plane-strain linear-elastic stiffness matrix $\bm{D}$. This also approximates the irreversibility condition through enforcing an increasing value for this history parameter. Since the force vector depends linearly on the phase field parameter, it is directly solved through $\bm{\upphi}^{t+\Delta t}=-\bm{K}_{\phi \phi}^{-1}\mathbf{f}_{\phi}+\bm{\upphi}^t$, using the tangent matrix:
\begin{equation}
    \bm{K}_{\phi\phi} = \int_{\Omega_\mathrm{s}} \frac{1}{\ell} \mathbf{N}_\phi^T\mathbf{N}_\phi + \ell \left(\bm{\nabla}\mathbf{N}_\phi\right)^T\bm{\nabla}\mathbf{N}_\phi  + 2 \left(1-k_\mathrm{0}\right)\mathcal{H}^{t+\Delta t} \mathbf{N}_\phi^T \mathbf{N}_\phi  \; \mathrm{d}{\Omega_\mathrm{s}}
\end{equation}

\subsubsection{Momentum balance sub-problem}
The second step is solving for the displacements through the momentum balance from \cref{eq:momentumbalance}. The discretised weak form is given by:
\begin{equation}
\begin{split}
    \mathbf{f}_\mathrm{u} = &\int_{\Omega_\mathrm{s}} \left(k_\mathrm{0}+(1-k_\mathrm{0})\left(1-\mathbf{N}_\phi \bm{\upphi}^{t+\Delta t}\right)^2\right)\bm{B}_\mathrm{u}^T \bm{D} \bm{B}_\mathrm{u} \mathbf{u}^{t+\Delta t}\;\mathrm{d}{\Omega_\mathrm{s}} \\ &- \int_\Gamma \left(k_\mathrm{0}+(1-k_\mathrm{0})\left(1-\mathbf{N}_\phi \bm{\upphi}^{t+\Delta t}\right)^2\right) \mathbf{N}_\mathrm{u}^T \bm{\uptau}_{\mathrm{ext}}\;\mathrm{d}\Gamma
\end{split}
\end{equation}
Since this equation is linear with regards to the nodal displacements $\mathbf{u}$, it is directly resolved through $\mathbf{u}^{t+\Delta t}=-\bm{K}_{\mathrm{u} \mathrm{u}}^{-1}\mathbf{f}_{\mathrm{u}}+\mathbf{u}^t$, using the tangent matrix:
\begin{equation}
    \bm{K}_{\mathrm{uu}} = \int_{\Omega_\mathrm{s}} \left(k_\mathrm{0}+(1-k_\mathrm{0})\left(1-\mathbf{N}_\phi \bm{\upphi}^{t+\Delta t}\right)^2\right)\bm{B}_\mathrm{u}^T \bm{D} \bm{B}_\mathrm{u} \; \mathrm{d}{\Omega_\mathrm{s}}
\end{equation}
Since the phase field is resolved first, and its updated value is used to compute the displacements, this step provides stresses and displacements that are compatible with the current state of the phase field. This is in contrast to a scheme where the displacements are determined first, and then used to update the phase field variable. As the electrochemical system resolved during the next step is strongly dependent on the hydrostatic stress gradient and the displacement field (via the crack opening height), this solution sequence was seen to be beneficial for the overall convergence. 

\subsubsection{Electrochemical sub-problem}
The last solution step resolves the electrochemical sub-problem. This is defined through the discretised weak form of the interstitial lattice hydrogen mass balance, \cref{eq:HydrogenMassBalance}, given by:
\begin{equation}
\begin{split}
    \mathbf{f}_\mathrm{L}^{t+\Delta t} = &\int_{\Omega_\mathrm{s}} \frac{1}{\Delta t}\left( 1 + \frac{N_\mathrm{T}/N_\mathrm{L} \exp(-\Delta g_\mathrm{b}/RT)}{\left(\mathbf{N}_\mathrm{L}\mathbf{C}_\mathrm{L}^{t+\Delta t}/N_\mathrm{L}+\exp(-\Delta g_\mathrm{b}/RT)\right)^2} \right) \mathbf{N}_\mathrm{L}^T \mathbf{N}_\mathrm{L} \left(\mathbf{C}_\mathrm{L}^{t+\Delta t}-\mathbf{C}_\mathrm{L}^t\right) \;\mathrm{d}{\Omega_\mathrm{s}} \\
    +& \int_{\Omega_\mathrm{s}} \frac{D_\mathrm{L}}{1-\mathbf{N}_\mathrm{L}\mathbf{C}_\mathrm{L}^{t+\Delta t}}\left(\bm{\nabla}\mathbf{N}_\mathrm{L}\right)^T\bm{\nabla}\mathbf{N}_\mathrm{L}\mathbf{C}_\mathrm{L}^{t+\Delta t}
    -\frac{D_\mathrm{L} \overline{V}_\mathrm{H}}{RT} \left(\bm{\nabla}\mathbf{N}_\mathrm{L}\right)^T \mathbf{N}_\mathrm{L}\mathbf{C}_\mathrm{L}^{t+\Delta t} \mathbf{B}_\mathrm{u}^*\mathbf{u}^{t+\Delta t} \;\mathrm{d}{\Omega_\mathrm{s}} \\ -& \int_\Gamma \mathbf{N}_\mathrm{L}^T J_{\mathrm{ext}}\;\mathrm{d}\Gamma + \sum_{\mathrm{nds_{\mathrm{s}}}} L_{\mathrm{ss}} 2 \left(\nu_\mathrm{A} - \nu_\mathrm{A}'\right) + \sum_{\mathrm{nds}_\Gamma} L_{\mathrm{eint}} \left(\nu_\mathrm{A} - \nu_\mathrm{A}'\right) 
    \end{split} \label{eq:weak_lattice}
\end{equation}
using the displacement to gradient of hydrostatic stress mapping matrix $\bm{\nabla}\sigma_h=\bm{B}_\mathrm{u}^*\mathbf{u}$. A lumped integration scheme is used for the last term, the reaction rates of the absorption reaction \citep{Hageman2022b}. More details relating to this lumped scheme, its impact on stability and its interaction with the distribution functions for the crack-contained electrolyte are given in \cref{sec:LumpedInt}. 

In addition to the interstitial lattice hydrogen mass balance, the surface adsorbed hydrogen mass balance, \cref{eq:electrolyte_surface}, ionic species mass balances, \cref{eq:electrolyte_nernstplanck}, and the electroneutrality condition, \cref{eq:electrolyte_neutrality}, are also resolved within this solution step. The weak form of this surface mass balance is given by:
\begin{equation}
\begin{split}
    \mathbf{f}_{\theta}^{t+\Delta t} = &\int_{\Omega_\mathrm{f}} 2 \frac{\beta_\mathrm{s} N_{\mathrm{ads}}}{\Delta t} \mathbf{N}_\theta^T\mathbf{N}_\theta \left(\bm{\uptheta}^{t+\Delta t}-\bm{\uptheta}^t\right) \; \mathrm{d}\Omega_{\mathrm{f}} + \int_{\Gamma} \frac{ N_{\mathrm{ads}}}{\Delta t} \mathbf{N}_\theta^T\mathbf{N}_\theta \left(\bm{\uptheta}^{t+\Delta t}-\bm{\uptheta}^t\right) \; \mathrm{d}\Gamma \\ - &\sum_{\mathrm{nds_{s}}} 2 L_{\mathrm{ss}} \left(\nu_{\mathrm{Va}}-\nu_{\mathrm{Va}}'-\nu_{\mathrm{Ha}}-2\nu_\mathrm{T} - \nu_\mathrm{A} + \nu_\mathrm{A}'+ \nu_{\mathrm{Vb}} - \nu_{\mathrm{Vb}}'-\nu_{\mathrm{Hb}}\right) \\
    -&\sum_{\mathrm{nds_{\Gamma}}} 2 L_{\mathrm{eint}} \left(\nu_{\mathrm{Va}}-\nu_{\mathrm{Va}}'-\nu_{\mathrm{Ha}}-2\nu_\mathrm{T} - \nu_\mathrm{A} + \nu_\mathrm{A}'+ \nu_{\mathrm{Vb}} - \nu_{\mathrm{Vb}}'-\nu_{\mathrm{Hb}}\right)
    \end{split} \label{eq:weak_surface}
\end{equation}
where the lumped integration is performed over the nodes within the solid domain, $\mathrm{nds}_\mathrm{s}$, and over the nodes at the metal-electrolyte interface, $\mathrm{nds}_\Gamma$. The ion concentration mass balances are given by:
\begin{equation}
\begin{split}
    \mathbf{f}_{\mathrm{c}\pi}^{t+\Delta t} = &\int_{\Omega_{\mathrm{s}}} \beta_\mathrm{c} \mathbf{N}_\mathrm{C}^T\mathbf{N}_\mathrm{C} \left(\mathbf{C}_\pi^{t+\Delta t}-\mathbf{C}_\pi^t \right) 
    + D_\pi \left(\bm{\nabla}\mathbf{N}_\mathrm{C}\right)^T \bm{R}^T\bm{\beta}_\mathrm{d}\bm{R}\bm{\nabla}\mathbf{N}_\mathrm{C}\mathbf{C}_\pi^{t+\Delta t} \;\mathrm{d}{\Omega_{\mathrm{s}}} \\ 
    +& \int_{\Omega_{\mathrm{s}}} \frac{z_\pi F D_\pi}{RT} \left(\bm{\nabla}\mathbf{N}_\mathrm{C}\right)^T \bm{R}^T \bm{\beta}_\mathrm{d} \bm{R} \mathbf{N}_\mathrm{C}\mathbf{C}_\pi^{t+\Delta t} \bm{\nabla} \mathbf{N}_\varphi \bm{\upvarphi}^{t+\Delta t} \; \mathrm{d}{\Omega_{\mathrm{s}}}\\
    +&\int_{\Omega_{\mathrm{e}}} \mathbf{N}_\mathrm{C}^T\mathbf{N}_\mathrm{C} \left(\mathbf{C}_\pi^{t+\Delta t}-\mathbf{C}_\pi^t \right) 
    + D_\pi \left(\bm{\nabla}\mathbf{N}_\mathrm{C}\right)^T \bm{\nabla}\mathbf{N}_\mathrm{C}\mathbf{C}_\pi^{t+\Delta t} \;\mathrm{d}{\Omega_{\mathrm{e}}} \\ 
    +& \int_{\Omega_{\mathrm{e}}} \frac{z_\pi F D_\pi}{RT} \left(\bm{\nabla}\mathbf{N}_\mathrm{C}\right)^T \mathbf{N}_\mathrm{C}\mathbf{C}_\pi^{t+\Delta t} \bm{\nabla} \mathbf{N}_\varphi \bm{\upvarphi}^{t+\Delta t} \; \mathrm{d}{\Omega_{\mathrm{e}}}\\
    +& \int_\Gamma J_{\mathrm{ext},\pi}\;\mathrm{d}\Gamma + \sum_{\mathrm{nds}_\mathrm{s}} L_{\mathrm{sv}} R_\pi + 2L_{\mathrm{ss}} \nu_\pi + \sum_{\mathrm{nds}_\mathrm{e}} L_{\mathrm{ev}} R_\pi + \sum_{\mathrm{nds}_\Gamma} L_{\mathrm{eint}} \nu_\pi \label{eq:weak_concentration}
\end{split}
\end{equation}
and the electroneutrality condition is given by:
\begin{equation}
    \mathbf{f}_{\varphi}^{t+\Delta t} = \int_{\Omega_\mathrm{s}} \sum_\pi \beta_\mathrm{c} z_\pi \mathbf{N}_\mathrm{\varphi}^T\mathbf{N}_\mathrm{C} \mathbf{C}_\pi^{t+\Delta t} \; \mathrm{d}\Omega_{\mathrm{s}} + \int_{\Omega_\mathrm{e}} \sum_\pi z_\pi \mathbf{N}_\mathrm{\varphi}^T\mathbf{N}_\mathrm{C} \mathbf{C}_\pi^{t+\Delta t} \; \mathrm{d}\Omega_\mathrm{e} \label{eq:weak_neutrality}
\end{equation}

Since this system of equations is nonlinear, a Newton-Raphson scheme is used within this step to solve \cref{eq:weak_lattice,eq:weak_surface,eq:weak_concentration,eq:weak_neutrality} concurrently. This scheme is defined as:
\begin{equation}
    \begin{bmatrix}
        \bm{K}_{\mathrm{LL}} & \bm{K}_{\mathrm{L}\theta} & 0 & 0\\
        \bm{K}_{\theta \mathrm{L}} & \bm{K}_{\theta \theta} & \bm{K}_{\theta \mathrm{C}} & \bm{K}_{\theta \varphi}\\
        0 & \bm{K}_{\mathrm{C}\theta} & \bm{K}_{\mathrm{CC}} & \bm{K}_{\mathrm{C}\varphi} \\
        0 & 0 & \bm{K}_{\varphi \mathrm{C}} & 0
    \end{bmatrix}
    \begin{bmatrix}
        \mathrm{d}\mathbf{C}_\mathrm{L} \\ \mathrm{d}\bm{\uptheta} \\ \mathrm{d}\mathbf{C}_\pi \\ \mathrm{d}\bm{\upvarphi}
    \end{bmatrix}
    =
    -\begin{bmatrix}
        \mathbf{f}_\mathrm{L}^{t+\Delta t} \\ \mathbf{f}_{\theta}^{t+\Delta t} \\ \mathbf{f}_{\mathrm{c}\pi}^{t+\Delta t} \\ \mathbf{f}_{\varphi}^{t+\Delta t}
    \end{bmatrix} \label{eq:NR}
\end{equation}
with the sub-matrices being given by:
\begin{flalign}
\begin{split}
    \bm{K}_{\mathrm{LL}} &= \\
    \int_{\Omega_\mathrm{s}} \frac{\mathbf{N}_\mathrm{L}^T \mathbf{N}_\mathrm{L}}{\Delta t}&\left( 1 + \frac{N_\mathrm{T}/N_\mathrm{L} \exp(-\Delta g_\mathrm{b}/RT)}{\left(\mathbf{N}_\mathrm{L}\mathbf{C}_\mathrm{L}^{t+\Delta t}/N_\mathrm{L}+\exp(-\Delta g_\mathrm{b}/RT)\right)^2} - \frac{2 N_\mathrm{T}/N_\mathrm{L}^2 \exp(-\Delta g_\mathrm{b}/RT)\; \mathbf{N}_\mathrm{L}\left(\mathbf{C}_\mathrm{L}^{t+\Delta t}-\mathbf{C}_\mathrm{L}^t\right)}{\left(\mathbf{N}_\mathrm{L}\mathbf{C}_\mathrm{L}^{t+\Delta t}/N_\mathrm{L}+\exp(-\Delta g_\mathrm{b}/RT)\right)^3} \right) \;\mathrm{d}\Omega_\mathrm{s} \\
    +& \int_{\Omega_{\mathrm{s}}} \frac{D_\mathrm{L}}{1-\mathbf{N}_\mathrm{L}\mathbf{C}_\mathrm{L}^{t+\Delta t}}\left(\bm{\nabla}\mathbf{N}_\mathrm{L}\right)^T\bm{\nabla}\mathbf{N}_\mathrm{L} + \frac{D_\mathrm{L}}{N_\mathrm{L}\left(1-\mathbf{N}_\mathrm{L}\mathbf{C}_\mathrm{L}^{t+\Delta t}\right)^2}\left(\bm{\nabla}\mathbf{N}_\mathrm{L}\right)^T\bm{\nabla}\mathbf{N}_\mathrm{L}\mathbf{C}_\mathrm{L}^{t+\Delta t} \mathbf{N}_\mathrm{L}\;\mathrm{d}\Omega_\mathrm{s} \\
    -& \int_{\Omega_\mathrm{s}} \frac{D_\mathrm{L} \overline{V}_\mathrm{H}}{RT} \left(\bm{\nabla}\mathbf{N}_\mathrm{L}\right)^T \mathbf{N}_\mathrm{L}\mathbf{B}_\mathrm{u}^*\mathbf{u}^{t+\Delta t} \;\mathrm{d}\Omega_\mathrm{s} \\
    +&\sum_{\mathrm{nds}_\mathrm{s}} 2 L_{\mathrm{ss}} \left(\frac{\partial \nu_\mathrm{A}}{\partial C_\mathrm{L}} - \frac{\partial \nu_\mathrm{A}'}{\partial C_\mathrm{L}}\right) \bm{I}_{\mathrm{LL}} 
    +\sum_{\mathrm{nds}_\Gamma} L_{\mathrm{eint}} \left(\frac{\partial \nu_\mathrm{A}}{\partial C_\mathrm{L}} - \frac{\partial \nu_\mathrm{A}'}{\partial C_\mathrm{L}}\right) \bm{I}_{\mathrm{LL}} 
\end{split} &&
\end{flalign}
\begin{flalign}
    \bm{K}_{\mathrm{L}\theta} = \sum_{\mathrm{nds}_\mathrm{s}} 2 L_{\mathrm{ss}} \left(\frac{\partial \nu_\mathrm{A}}{\partial \theta} - \frac{\partial \nu_\mathrm{A}'}{\partial \theta}\right) \bm{I}_{\mathrm{L}\theta} + \sum_{\mathrm{nds}_\Gamma} L_{\mathrm{eint}} \left(\frac{\partial \nu_\mathrm{A}}{\partial \theta} - \frac{\partial \nu_\mathrm{A}'}{\partial \theta}\right) \bm{I}_{\mathrm{L}\theta} &&
\end{flalign}
\begin{flalign}
    \bm{K}_{\theta \mathrm{L}} = \sum_{\mathrm{nds}_\mathrm{s}} 2 L_{\mathrm{ss}} \left( \frac{\partial \nu_\mathrm{A}}{\partial C_\mathrm{L}} - \frac{\partial \nu_\mathrm{A}'}{\partial C_\mathrm{L}} \right) \bm{I}_{\theta \mathrm{L}} + \sum_{\mathrm{nds}_\Gamma} L_{\mathrm{eint}} \left( \frac{\partial \nu_\mathrm{A}}{\partial C_\mathrm{L}} - \frac{\partial \nu_\mathrm{A}'}{\partial C_\mathrm{L}} \right) \bm{I}_{\theta \mathrm{L}}&&
\end{flalign}
\begin{flalign}
    \begin{split}\bm{K}_{\theta\theta} &= \int_{\Omega_\mathrm{s}} 2 \frac{\left(\beta_\mathrm{s}+\epsilon\right) N_{\mathrm{ads}}}{\Delta t} \mathbf{N}_\theta^T\mathbf{N}_\theta \; \mathrm{d}\Omega_\mathrm{s} +  \int_{\Gamma} \frac{N_{\mathrm{ads}}}{\Delta t} \mathbf{N}_\theta^T\mathbf{N}_\theta \; \mathrm{d}\Gamma \\ & -\sum_{\mathrm{nds}_\mathrm{s}} 2 L_{\mathrm{ss}} \left(\frac{\partial \nu_{\mathrm{Va}}}{\partial \theta_{\mathrm{ads}}}-\frac{\partial \nu_{\mathrm{Va}}'}{\partial \theta_{\mathrm{ads}}}-\frac{\partial\nu_{\mathrm{Ha}}}{\partial \theta_{\mathrm{ads}}}-2\frac{\partial \nu_\mathrm{T}}{\partial \theta_{\mathrm{ads}}} - \frac{\partial\nu_\mathrm{A}}{\partial \theta_{\mathrm{ads}}} + \frac{\partial\nu_\mathrm{A}'}{\partial \theta_{\mathrm{ads}}}+ \frac{\partial\nu_{\mathrm{Vb}}}{\partial \theta_{\mathrm{ads}}} - \frac{\partial\nu_{\mathrm{Vb}}'}{\partial \theta_{\mathrm{ads}}}-\frac{\partial\nu_{\mathrm{Hb}}}{\partial \theta_{\mathrm{ads}}} \right)\bm{I}_{\theta\theta} \\
     &-\sum_{\mathrm{nds}_\Gamma} L_{\mathrm{eint}} \left(\frac{\partial \nu_{\mathrm{Va}}}{\partial \theta_{\mathrm{ads}}}-\frac{\partial \nu_{\mathrm{Va}}'}{\partial \theta_{\mathrm{ads}}}-\frac{\partial\nu_{\mathrm{Ha}}}{\partial \theta_{\mathrm{ads}}}-2\frac{\partial \nu_\mathrm{T}}{\partial \theta_{\mathrm{ads}}} - \frac{\partial\nu_\mathrm{A}}{\partial \theta_{\mathrm{ads}}} + \frac{\partial\nu_\mathrm{A}'}{\partial \theta_{\mathrm{ads}}}+ \frac{\partial\nu_{\mathrm{Vb}}}{\partial \theta_{\mathrm{ads}}} - \frac{\partial\nu_{\mathrm{Vb}}'}{\partial \theta_{\mathrm{ads}}}-\frac{\partial\nu_{\mathrm{Hb}}}{\partial \theta_{\mathrm{ads}}} \right)\bm{I}_{\theta\theta}
    \end{split} &&
\end{flalign}
\begin{flalign}
\begin{split}
    \bm{K}_{\theta \mathrm{C}} = &- \sum_{\mathrm{nds}_\mathrm{s}} 2 L_{\mathrm{ss}} \left(\frac{\partial\nu_{\mathrm{Va}}}{\partial C_\pi}-\frac{\partial \nu_{\mathrm{Va}}'}{\partial C_\pi}-\frac{\partial \nu_{\mathrm{Ha}}}{\partial C_\pi} + \frac{\partial \nu_{\mathrm{Vb}}}{\partial C_\pi} - \frac{\partial \nu_{\mathrm{Vb}}'}{\partial C_\pi}-\frac{\partial \nu_{\mathrm{Hb}}}{\partial C_\pi} \right)\bm{I}_{\theta \mathrm{C}} \\ 
    &- \sum_{\mathrm{nds}_\Gamma} L_{\mathrm{eint}} \left(\frac{\partial\nu_{\mathrm{Va}}}{\partial C_\pi}-\frac{\partial \nu_{\mathrm{Va}}'}{\partial C_\pi}-\frac{\partial \nu_{\mathrm{Ha}}}{\partial C_\pi} + \frac{\partial \nu_{\mathrm{Vb}}}{\partial C_\pi} - \frac{\partial \nu_{\mathrm{Vb}}'}{\partial C_\pi}-\frac{\partial \nu_{\mathrm{Hb}}}{\partial C_\pi} \right)\bm{I}_{\theta \mathrm{C}}
\end{split} &&
\end{flalign}
\begin{flalign}
\begin{split}
    \bm{K}_{\theta \varphi} = &- \sum_{\mathrm{nds}_\mathrm{s}} 2 L_{\mathrm{ss}} \left(\frac{\partial\nu_{\mathrm{Va}}}{\partial \varphi}-\frac{\partial \nu_{\mathrm{Va}}'}{\partial \varphi}-\frac{\partial \nu_{\mathrm{Ha}}}{\partial \varphi} + \frac{\partial \nu_{\mathrm{Vb}}}{\partial \varphi} - \frac{\partial \nu_{\mathrm{Vb}}'}{\partial \varphi}-\frac{\partial \nu_{\mathrm{Hb}}}{\partial \varphi} \right) \bm{I}_{\theta\varphi}\\
    &- \sum_{\mathrm{nds}_\Gamma} L_{\mathrm{eint}} \left(\frac{\partial\nu_{\mathrm{Va}}}{\partial \varphi}-\frac{\partial \nu_{\mathrm{Va}}'}{\partial \varphi}-\frac{\partial \nu_{\mathrm{Ha}}}{\partial \varphi} + \frac{\partial \nu_{\mathrm{Vb}}}{\partial \varphi} - \frac{\partial \nu_{\mathrm{Vb}}'}{\partial \varphi}-\frac{\partial \nu_{\mathrm{Hb}}}{\partial \varphi} \right) \bm{I}_{\theta\varphi}
\end{split} &&
\end{flalign}
\begin{flalign}
    \bm{K}_{\mathrm{C} \theta} = \sum_{\mathrm{nds}_\mathrm{s}} 2L_{\mathrm{ss}} \frac{\partial \nu_\pi}{\partial \theta} \bm{I}_\theta  + \sum_{\mathrm{nds}_\Gamma} L_{\mathrm{eint}} \frac{\partial \nu_\pi}{\partial \theta} \bm{I}_\theta &&
\end{flalign}
\begin{flalign}
    \begin{split} \bm{K}_{\mathrm{CC}} &= \int_{\Omega_\mathrm{s}} \left(\epsilon+\beta_\mathrm{c}\right) \mathbf{N}_\mathrm{C}^T\mathbf{N}_\mathrm{C} 
    + D_\pi \left(\bm{\nabla}\mathbf{N}_\mathrm{C}\right)^T \left(\bm{R}^T\bm{\beta}_\mathrm{d}\bm{R}+\epsilon\bm{I}\right)\bm{\nabla}\mathbf{N}_\mathrm{C} \;\mathrm{d}\Omega_\mathrm{s} \\ 
    &+ \int_{\Omega_\mathrm{s}} \frac{z_\pi F D_\pi}{RT} \left(\bm{\nabla}\mathbf{N}_\mathrm{C}\right)^T \left(\bm{R}^T \bm{\beta}_\mathrm{d} \bm{R}+\epsilon \bm{I}\right) \mathbf{N}_\mathrm{C} \bm{\nabla} \mathbf{N}_\varphi \bm{\upvarphi}^{t+\Delta t} \; \mathrm{d}\Omega_\mathrm{s} \\
    &+ \int_{\Omega_\mathrm{e}} \mathbf{N}_\mathrm{C}^T\mathbf{N}_\mathrm{C} 
    + D_\pi \left(\bm{\nabla}\mathbf{N}_\mathrm{C}\right)^T \bm{\nabla}\mathbf{N}_\mathrm{C} \;\mathrm{d}\Omega_\mathrm{e} \\ 
    &+ \int_{\Omega_\mathrm{e}} \frac{z_\pi F D_\pi}{RT} \left(\bm{\nabla}\mathbf{N}_\mathrm{C}\right)^T \mathbf{N}_\mathrm{C} \bm{\nabla} \mathbf{N}_\varphi \bm{\upvarphi}^{t+\Delta t} \; \mathrm{d}\Omega_\mathrm{e} \\
    &+ \sum_{\mathrm{nds}_\mathrm{s}} L_{\mathrm{sv}} \frac{\partial R_\pi}{\partial C_{\pi}} + 2L_{\mathrm{ss}} \frac{\partial \nu_\pi}{\partial C_\pi} \bm{I}_{\mathrm{CC}} + \sum_{\mathrm{nds}_e} L_{\mathrm{ev}} \frac{\partial R_\pi}{\partial C_{\pi}} \bm{I}_{\mathrm{CC}} + \sum_{\mathrm{nds}_\Gamma} L_{\mathrm{ls}} \frac{\partial \nu_\pi}{\partial C_\pi} \bm{I}_{\mathrm{CC}} \end{split} &&
\end{flalign}
\begin{flalign}
\begin{split}
    \bm{K}_{\mathrm{C} \varphi} = &\int_{\Omega_\mathrm{s}} \frac{z_\pi F D_\pi}{RT} \left(\bm{\nabla}\mathbf{N}_\mathrm{C}\right)^T \left(\bm{R}^T \bm{\beta}_\mathrm{d} \bm{R}+\epsilon \bm{I}\right) \mathbf{N}_\mathrm{C}\mathbf{C}_\pi^{t+\Delta t} \bm{\nabla} \mathbf{N}_\varphi \; \mathrm{d}\Omega_\mathrm{s} + \sum_{\mathrm{nds}_\mathrm{s}} 2L_{\mathrm{ss}} \frac{\partial \nu_\pi}{\partial \varphi} \\
    +& \int_{\Omega_\mathrm{e}} \frac{z_\pi F D_\pi}{RT} \left(\bm{\nabla}\mathbf{N}_\mathrm{C}\right)^T \mathbf{N}_\mathrm{C}\mathbf{C}_\pi^{t+\Delta t} \bm{\nabla} \mathbf{N}_\varphi \; \mathrm{d}\Omega_\mathrm{e} + \sum_{\mathrm{nds}_\gamma} L_{\mathrm{eint}} \frac{\partial \nu_\pi}{\partial \varphi}
\end{split} &&
\end{flalign}
\begin{flalign}
    \bm{K}_{\varphi \mathrm{C}} = \int_{\Omega_{\mathrm{s}}} \sum_\pi \left(\beta_\mathrm{c}+\epsilon\right) z_\pi \mathbf{N}_\mathrm{\varphi}^T\mathbf{N}_\mathrm{C} \; \mathrm{d}\Omega_\mathrm{s} + \int_{\Omega_\mathrm{e}} \sum_\pi z_\pi \mathbf{N}_\mathrm{\varphi}^T\mathbf{N}_\mathrm{C} \; \mathrm{d}\Omega_\mathrm{e}&&
\end{flalign}
These tangent matrices use the allocation matrices $\bm{I}_{xy}$, to allocate the lumped integration terms to the correct locations within the stiffness matrix, as given by the set of degrees of freedom ($x,y$). The matrices related to capacity and diffusion terms also contain an offset $\epsilon$, whose presence is explained in \cref{sec:fix_unconstrained_dofs}. The system from \cref{eq:NR} is iterated until a converged solution is achieved, using an energy based criterion:
\begin{equation}
    E_{it} = E_{it}^*/E_\mathrm{0}<10^{-6} \qquad   \text{with} \,\,\,\,\,\,\, E^*_{it} = [\mathbf{f}_\mathrm{L}^T\;\mathbf{f}_\theta^T\;\mathbf{f}_{\mathrm{c}\pi}^T\;\mathbf{f}_\varphi^T ]_{it}[\mathrm{d} \mathbf{C}_\mathrm{L};\;\mathrm{d}\mathbf{\uptheta};\;\mathrm{d}\mathbf{C}_{\pi};\;\mathrm{d}\mathbf{\upvarphi}]_{it}
\end{equation}
Upon convergence, the errors within the phase field evolution and momentum balance are calculated and compared to the criterion $E_{it}^*<10^{-6}$. If this is fulfilled, the simulation proceeds to the next time increment. If the error is exceeded, another staggered iteration is performed solving the phase field, displacements and electrochemical systems. 

\subsection{Stabilising effect of lumped integration}
\label{sec:LumpedInt}
One issue when simulating electro-chemical systems using finite elements is the large range of reaction rates present. As these rates depend strongly on the environment, often varying by many orders of magnitude at different locations within the same simulation, it is often not feasible to enforce these reactions to be \textit{a priori} in instant equilibrium. Furthermore, enforcing a direct equilibrium between reactions is often accomplished by eliminating them from the governing equations of the system, greatly complicating the addition of other reactions involving the eliminated species. However, as a result of the (potentially) very high reaction rates, a stiff system of differential equations is created. Solving this system of equations poses difficulties, with ill-conditioned tangent matrices and results that often exhibit non-physical oscillations.

\begin{figure}
     \centering
     \begin{subfigure}[b]{0.49\textwidth}
         \centering
         \includegraphics[clip, trim={0 0 0 50}]{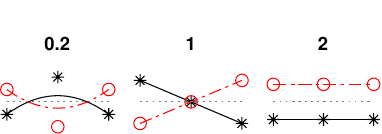}
         \caption{Gauss integration}
         \label{fig:ExampleGauss}
     \end{subfigure}
     \begin{subfigure}[b]{0.49\textwidth}
         \centering
         \includegraphics[clip, trim={0 0 0 50}]{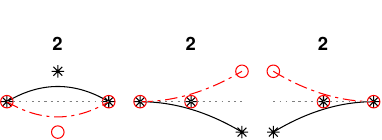}
         \caption{Lumped integration}
         \label{fig:ExampleLumped}
     \end{subfigure}
        \caption{Non-zero eigenmodes obtained using Gauss (a) and lumped (b) integration evaluated for the matrices of \cref{eq:KMats}. Black lines and circles are used to indicate the change in surface occupancy due to the adsorption reaction, while red dashed lines and stars indicate the changes in interstitial lattice hydrogen concentration.}
        \label{fig:ExampleGaussLumped}
\end{figure}

One manner in which these difficulties can be tackled is by using lumped integration for the problematic reaction terms \citep{Hageman2022b}. Originally developed to resolve issues with traction oscillations due to contact conditions when using interface elements \citep{Schellekens1993}, lumped integration performs the integration of transfer terms (such as electro-chemical reactions) on a node-by-node basis in a consistent manner. For instance, considering the hydrogen absorption term within \cref{eq:weak_surface,eq:weak_lattice}:
\begin{alignat}{2}
    \mathbf{f}_\mathbf{L}^{\mathrm{abs}} &= &&\int_{\Omega_s} 2 \beta_s \mathbf{N}_\mathrm{C}^T \left(k_\mathrm{A}\left(N_L - \mathbf{N}_\mathrm{L} \mathbf{C}_\mathrm{L}\right) \mathbf{N}_\theta \mathbf{\uptheta} - k_A'\left(1-\mathbf{N}_\theta \mathbf{\uptheta} \right)\mathbf{N}_L \mathbf{C}_L\right)\;\mathrm{d}\Omega_\mathrm{s} \label{eq:ExampleLumped1}
    \\
    \mathbf{f}_\theta^{\mathrm{abs}} &= -&&\int_{\Omega_s} 2 \beta_s \mathbf{N}_\theta^T \left(k_\mathrm{A}\left(N_L - \mathbf{N}_\mathrm{L} \mathbf{C}_\mathrm{L}\right) \mathbf{N}_\theta \mathbf{\uptheta} - k_A'\left(1-\mathbf{N}_\theta \mathbf{\uptheta}\right) \mathbf{N}_L \mathbf{C}_L\right)\;\mathrm{d}\Omega_\mathrm{s} \label{eq:ExampleLumped2}
\end{alignat}
Using a standard Gauss integration scheme, these integrals are directly evaluated through a sum over their integration points. In contrast, when using a lumped integration scheme, consistent weights for these surface reactions are first determined as:
\begin{equation}
    \mathbf{L}_{ss} = \int_{\Omega_\mathrm{s}} \beta_\mathrm{s} \mathbf{N}^T\;\mathrm{d}\Omega_\mathrm{s} = \sum_{\mathrm{el}_\mathrm{s}} \sum_{\mathrm{ip}} w_{\mathrm{ip}} \beta_{\mathrm{s}}(\phi_\mathrm{ip}) \mathbf{N}^T
\end{equation}
where the lumped integration weights associated with each node are obtained using a standard Gauss integration scheme, as a sum over elements and integration points. Having calculated consistent weights for each node, the lumped integration of \cref{eq:ExampleLumped1,eq:ExampleLumped2} is performed as a sum over all nodes:
\begin{alignat}{2}
    \mathbf{f}_\mathrm{L}^{\mathrm{abs}} & = && \sum_{n=\mathrm{nds}_\mathrm{s}} 2\mathbf{L}_{ss}^{n} \left(k_\mathrm{A}\left(N_L - \mathbf{C}_\mathrm{L}^n\right) \mathbf{\uptheta}^n - k_A'\left(1-\mathbf{\uptheta}^n\right) \mathbf{C}_L^n\right) \mathbf{i}_\mathrm{L}^n\\
    \mathbf{f}_\mathrm{\theta}^{\mathrm{abs}} & = && \sum_{n=\mathrm{nds}_\mathrm{s}} 2\mathbf{L}_{ss}^{n} \left(k_\mathrm{A}\left(N_L - \mathbf{C}_\mathrm{L}^n\right) \mathbf{\uptheta}^n - k_A'\left(1-\mathbf{\uptheta}^n\right) \mathbf{C}_L^n\right) \mathbf{i}_\theta^n
\end{alignat}
using the superscript $n$ to indicate the nodal values, and $\mathbf{i}_\mathrm{L}^n$ to denote the row of the force vector corresponding to the correct degree of freedom and  node $n$. To illustrate the effect of this lumped integration scheme on the systems tangent matrices, we shall calculate these using Gauss and lumped integration (setting $k_\mathrm{A}=k_\mathrm{A}'=3$, $N_\mathrm{L}=1$, $\beta_\mathrm{s}=1$). For brevity, quadratic line elements are used here, while quadratic quad and triangular elements are used within the actual implementation. Setting $\mathbf{C}_\mathrm{L}=\mathbf{\uptheta}=\mathbf{0}$ results in the following tangent matrices:
\begin{equation}
    \bm{K}_{Gauss} = \begin{bmatrix}
        0.6 & 0.3 & 0.1 & -0.6 & -0.3 & -0.1\\
        0.3 & 0.4 & 0.3 & -0.3 & -0.4 & -0.3\\
        0.1 & 0.3 & 0.6 & -0.1 & -0.3 & -0.6\\
        -0.6 & -0.3 & -0.1 & 0.6 & 0.3 & 0.1\\
        -0.3 & -0.4 & -0.3 & 0.3 & 0.4 & 0.3\\
        -0.1 & -0.3 & -0.6 & 0.1 & 0.3 & 0.6
    \end{bmatrix} \quad     \bm{K}_{Lumped} = \begin{bmatrix}
        1 & 0 & 0 & -1 & 0 & 0\\
        0 & 1 & 0 & 0 & -1 & 0\\
        0 & 0 & 1 & 0 & 0 & -1\\
        -1 & 0 & 0 & 1 & 0 & 0\\
        0 & -1 & 0 & 0 & 1 & 0\\
        0 & 0 & -1 & 0 & 0 & 1
    \end{bmatrix} \label{eq:KMats}
\end{equation}
As is evident from the first three elements of the two matrices, the tangent matrix obtained using Gauss integration allows interactions between neighbouring nodes, transferring species to different locations via chemical reactions. In contrast, the lumped matrix solely allows reactions to occur between degrees of freedom co-located in the same node. The effect of this is also seen by looking at the eigenmodes described by these matrices, with the non-zero eigenmodes shown in \cref{fig:ExampleGaussLumped}. The two lowest eigenmodes when using a Gauss scheme correspond to transfer of chemical species between neighbouring nodes without any change in the total amount of these species. It is only with the addition of the third and highest eigenmode that chemical reactions become possible. In contrast, the lumped tangent matrix obtains three equal eigenmodes, corresponding to reactions between degrees of freedom in the same node. 

In a similar manner, the nodal integration weights for the volume reaction terms used within the phase field electrolyte description are given by:
\begin{equation}
    \mathbf{L}_{\mathrm{sv}} = \int_{\Omega_\mathrm{s}} \beta_\mathrm{c} \mathbf{N}_\mathrm{L} \; \mathrm{d}{\Omega_\mathrm{s}}
\end{equation}
and the lumped weights used for the free electrolyte and metal-electrolyte interface reactions are given by:
\begin{equation}
    \mathbf{L}_{\mathrm{ev}} = \int_{\Omega_{\mathrm{e}}} \mathbf{N}_\mathrm{C} \; \mathrm{d}{\Omega_{\mathrm{e}}}
\end{equation}
\begin{equation}
    \mathbf{L}_{\mathrm{eint}} = \int_{\Gamma} \mathbf{N}_\mathrm{L} \; \mathrm{d}{\Gamma}
\end{equation}
This lumped integration is applied to all reaction terms. While some reactions are not dominant and would not cause any numerical difficulties, this is highly dependent on the local conditions. Furthermore, as the lumped integration scheme is consistent with the weak form, using a lumped integration instead of a Gauss integration scheme has little or no effect on the obtained solution \cite{Hageman2022b}.

\subsection{Prevention of ill-constrained degrees of freedom}
\label{sec:fix_unconstrained_dofs}

One issue while solving \cref{eq:electrolyte_nernstplanck} or \cref{eq:weak_concentration} is that, for elements where $\phi \approx 0$, the multiplication with either the phase field parameter or the surface distribution function can cause unconstrained degrees of freedom. To remedy this, while not altering the obtained solution for the well-defined degrees of freedom, inconsistent tangent matrices are constructed using altered distribution functions. These distribution functions include a small offset $\epsilon$ to prevent the system from becoming ill-defined, for instance defining the contribution to the internal force vector of the ion capacity term as:
\begin{equation}
    \mathbf{f}_{\mathrm{c}\pi}^{\mathrm{capacity}} = \int_{\Omega_\mathrm{s}}  h \left( \frac{1}{2\ell}\phi^2+\frac{\ell}{2}\left|\bm{\nabla} \phi \right|^2 \right) \frac{1}{\Delta t}\mathbf{N}_\mathrm{C}'\mathbf{N}_\mathrm{C} \left(\mathbf{C}_{\pi}^{t+\Delta t} - \mathbf{C}_{\pi}^t\right)\;\mathrm{d}\Omega_\mathrm{s}
\end{equation}
while the tangential term contributing to the tangential matrix of the system is defined as:
\begin{equation}
    \bm{K}_{\pi\pi}^{\mathrm{capacity}} = \int_{\Omega_\mathrm{s}}  \left(\epsilon + h \left( \frac{1}{2\ell}\phi^2+\frac{\ell}{2}\left|\bm{\nabla} \phi \right|^2 \right)\right) \frac{1}{\Delta t} \mathbf{N}_\mathrm{C}'\mathbf{N}_\mathrm{C} \;\mathrm{d}\Omega_\mathrm{s}
\end{equation}
This allows the electro-chemical degrees of freedom to remain constrained. Since the offset is solely introduced in the tangent matrix, it does not alter the converged solution state, instead only altering the rate at which this converged state is obtained. Small values of $\epsilon$ prevent the system from becoming ill-constrained, but significantly alter the conditioning number of the matrix due to the many orders of magnitude difference between the terms within and outside the cracks. Increasing the value of $\epsilon$ improves this matrix conditioning and enhances the stability of the solver, at the cost of requiring more iterations to obtain a well-converged solution. A value of $\epsilon = 10^{-12}$ is used throughout this paper.

\subsection{Initialisation of phase field parameter and history field}
For the initialisation at the start of simulations, we set the displacements, electrolyte potential, surface coverage, and interstitial lattice, $\mathrm{Fe}^{2+}$ and $\mathrm{FeOH}^+$ concentrations to zero. The $\mathrm{H}^+$ and $\mathrm{OH}^-$ concentrations are initialised as equal to the imposed boundary $\mathrm{pH}$, and the $\mathrm{Na}^+$ and $\mathrm{Cl}^-$ concentrations are set equal to the boundary values. At the start of the simulations, an initial fracture is assumed to be already present. While it is common for this fracture to be represented geometrically \citep{Miehe2010, Miehe2015, Duda2018, Gerasimov2019, Kristensen2021, Quinteros2022, Egger2019}, we choose to include it by setting initial values for the phase field and history variable based on the distance $\text{d}x$ from the preferred initial fracture:
\begin{equation}
    \phi^{\mathrm{init}} = \exp\left(-\left|\text{d}x\right|/\ell\right) 
    \qquad \qquad 
    \mathcal{H}^{\mathrm{init}} = \frac{1/\ell \; \; \mathbf{N}_\phi \bm{\upphi}^{\mathrm{init}} + \ell \left(\bm{\nabla}\mathbf{N}_\phi \bm{\upphi}^{\mathrm{init}}\right)^T\left(\bm{\nabla}\mathbf{N}_\phi \bm{\upphi}^{\mathrm{init}}\right)}{k_\mathrm{0}-2(1-k_{\mathrm{0}})\left(1-\mathbf{N}_{\phi}\bm{\upphi}^{\mathrm{init}}\right)} \label{eq:initialization}
\end{equation}
which is based on the one-dimensional solution of the phase field function. While this does not provide an exact solution for higher dimensional cases, it is sufficient to trigger the localisation of the phase field required to obtain a fracture consistent with the preferred initial crack after the first time increment. The main advantage of including the initial fracture through the phase field is the automatic inclusion of the electrolyte within the initial crack, whereas had this been represented geometrically, an additional set of equations would have been required.

\section{Results}
\label{sec:results}

\begin{table}
 \caption{Material parameters relevant to the metal domain.}
\label{tab:propertiesMetal}
        \centering
\begin{tabular}{ |l l||l|  }
 \hline
  Parameter & & Value\\
 \hline
     Young's Modulus & $E$ & $200\;\mathrm{GPa}$\\
     Poisson ratio & $\nu$ & $0.3$\\
     Fracture release energy & $G_{\mathrm{c0}}$ & $2\cdot10^3\;\mathrm{J}/\mathrm{m}^2$\\
     Residual stiffness factor & $k_\mathrm{0}$ & $10^{-10}$\\
     Degradation factor & $\chi$ & $0.9$\\
     Grain boundary binding energy & $\Delta g_\mathrm{b}$ & $30\;\mathrm{kJ}/\mathrm{mol}$\\
     Grain boundary site concentration & $N_\mathrm{T}$ & $10^2\;\mathrm{mol}/\mathrm{m}^3$\\
     Interstitial lattice site concentration & $N_\mathrm{L}$ & $10^6\;\mathrm{mol}/\mathrm{m}^3$\\
     Hydrogen diffusivity & $D_\mathrm{L}$ & $10^{-9}\;\mathrm{m}/\mathrm{s}$\\
     Reference temperature & $T$ & $20\;^\circ \mathrm{C}$\\
\hline
\end{tabular}
\end{table}

\begin{table}
 \caption{Parameters relevant to the electrolyte domain.}
\label{tab:propertiesElectrolyte}
        \centering
\begin{tabular}{ |l l||l|  }
 \hline
  Parameter & & Value\\
 \hline
    $\mathrm{H}^+$ diffusivity & $D_{\mathrm{H}^+}$ & $9.3\cdot10^{-9}\;\mathrm{m}/\mathrm{s}$\\
    $\mathrm{OH}^-$ diffusivity & $D_{\mathrm{H}^-}$ & $5.3\cdot10^{-9}\;\mathrm{m}/\mathrm{s}$\\
    $\mathrm{Na}^+$ diffusivity & $D_{\mathrm{Na}^+}$ & $1.3\cdot10^{-9}\;\mathrm{m}/\mathrm{s}$\\
    $\mathrm{Cl}^-$ diffusivity & $D_{\mathrm{Cl}^-}$ & $2\cdot10^{-9}\;\mathrm{m}/\mathrm{s}$\\
    $\mathrm{Fe}^{2+}$ diffusivity & $D_{\mathrm{Fe}^{2+}}$ & $1.4\cdot10^{-9}\;\mathrm{m}/\mathrm{s}$\\
    $\mathrm{FeOH}^+$ diffusivity & $D_{\mathrm{FeOH}^+}$ & $1\cdot10^{-9}\;\mathrm{m}/\mathrm{s}$\\
\hline
    Initial concentration $\mathrm{H}^+$ & $C_{0\mathrm{H}^+}$ & $10^{-2}\;\mathrm{mol}/\mathrm{m^3}$\\
    Initial concentration $\mathrm{OH}^-$ & $C_{0\mathrm{OH}^-}$ & $10^{-6}\;\mathrm{mol}/\mathrm{m^3}$\\
    Initial concentration $\mathrm{Na}^+$ & $C_{0\mathrm{Na}^+}$ & $600\;\mathrm{mol}/\mathrm{m^3}$\\
    Initial concentration $\mathrm{Cl}^-$ & $C_{0\mathrm{Cl}^-}$ & $\approx600\;\mathrm{mol}/\mathrm{m^3}$ \tablefootnote{Value corrected for electroneutrality condition}\\
    Initial concentration $\mathrm{Fe}^{2+}$ & $C_{0\mathrm{Fe}^{2+}}$ & $0\;\mathrm{mol}/\mathrm{m^3}$\\
    Initial concentration $\mathrm{FeOH}^+$ & $C_{0\mathrm{FeOH}^+}$ & $0\;\mathrm{mol}/\mathrm{m^3}$\\
\hline
    Boundary potential & $\varphi_\mathrm{0}$ & $0\;\mathrm{V}_{\mathrm{SHE}}$\\
    Metal Potential & $E_\mathrm{m}$ & $0\;\mathrm{V}_{\mathrm{SHE}}$\\
\hline
\end{tabular}
\end{table}

\begin{table}
 \caption{Parameters relevant to the electrochemical reactions.} \label{tab:propertiesReactions}
        \centering
\begin{tabular}{ |l|l|l|l|l|  }
 \hline
  Reaction & $k$ & $k'$ & $\alpha$ & $E_{\mathrm{eq}}$\\
  \hline 
 $\nu_{\mathrm{Va}}$ & $1\cdot10^{-4}\; \mathrm{m}/\mathrm{s}$ & $1\cdot10^{-10}\;\mathrm{mol/(m}^2\mathrm{s)}$ & $0.5$ & $0\;\mathrm{V}_{\mathrm{SHE}}$ \\
 $\nu_{\mathrm{Ha}}$ & $1\cdot10^{-10} \;\mathrm{m/s}\;\;$ & $0 \;\mathrm{mol/(m}^2\mathrm{Pa\; s)}$ & $0.3$ & $0\;\mathrm{V}_{\mathrm{SHE}}$\\ 
 $\nu_\mathrm{T}$ & $1\cdot10^{-6} \;\mathrm{mol/(m}^2\mathrm{s)}$ & $0 \;\mathrm{mol/(m}^2\mathrm{s \;Pa})$ & $-$ & $-$ \\ 
 $\nu_\mathrm{A}$ & $1\cdot10^1 \;\mathrm{m/s}$ & $7\cdot10^5 \;\mathrm{m/s}$ & $-$ & $-$ \\ 
 $\nu_{\mathrm{Vb}}$ & $1\cdot10^{-8} \; \mathrm{mol/(m}^2\mathrm{s})$ & $1\cdot10^{-13} \;\mathrm{m/s}$ & $0.5$ & $0\;\mathrm{V}_{\mathrm{SHE}}$ \\
 $\nu_{\mathrm{Hb}}$ & $1\cdot10^{-10} \;\mathrm{mol/(m}^2\mathrm{s)}$ & $0 \;\mathrm{m/(Pa \;s)}$ & $0.3$ & $0\;\mathrm{V}_{\mathrm{SHE}}$ \\
 $\nu_{\mathrm{c}}$ & $1.5\cdot10^{-10}\;\mathrm{mol}/\mathrm{(m}^2\mathrm{s)}$ & $1.5\cdot10^{-10}\;\mathrm{m}/\mathrm{s}$ & $0.5$ & $-0.4\;\mathrm{V}_{\mathrm{SHE}}$ \\
 \hline
$k_{\mathrm{fe}}$ & $0.1\;\mathrm{s}$ & $10^{-3}\;\mathrm{m}^3/(\mathrm{mol}\;\mathrm{s})$ &  &  \\
$k_{\mathrm{feoh}}$ & $10^{-3} \;\mathrm{s}^{-1}$ & & &  \\
$k_{\mathrm{eq}}$ & $10^6\;\mathrm{m}^3/(\mathrm{mol}\;\mathrm{s})$ & & & \\
\hline
\end{tabular}
\end{table}

The accuracy and applicability of the described model is demonstrated through a set of case studies. First, we study electrolyte behaviour in a stationary crack, so as to benchmark our physics-based treatment of electrolytes within cracks against discrete simulations and other existing phase field-based models (see \cref{sec:case1}). Then, in \cref{sec:case2}, we simulate the propagation of cracks exposed to a hydrogen-containing electrolyte, to showcase the main predictive capabilities of the model. Finally, in \cref{sec:case3}, we extend our analysis to a case study containing both free and crack-contained electrolytes. These three case studies all use the metal properties given in \cref{tab:propertiesMetal}, the electrolyte properties given in \cref{tab:propertiesElectrolyte}, and the reaction rate constants listed in \cref{tab:propertiesReactions}, with this set of properties corresponding to an iron-based metal in contact with seawater.

\subsection{Benchmark case study: handling electrolyte-containing cracks}
\label{sec:case1}
\begin{figure}
     \centering
     \begin{subfigure}[b]{0.49\textwidth}
         \centering
         \includegraphics{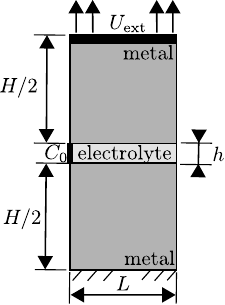}
         \caption{}
         \label{fig:Case1_DomainsA}
     \end{subfigure}
     \begin{subfigure}[b]{0.49\textwidth}
         \centering
         \includegraphics{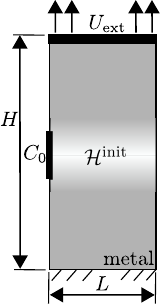}
         \caption{}
         \label{fig:Case1_DomainsB}
     \end{subfigure}
        \caption{Physical domain considered and directly simulated in the reference simulations (a), and the domain and boundary conditions relevant to the phase field representation (b).}
        \label{fig:Case1_Domains}
\end{figure}

We first investigate the capabilities of the physics-based model presented in Section \ref{sec:ModelPhysicsBased}. To this end, we consider a boundary value problem containing two metallic regions divided by an electrolyte. This benchmark geometry allows us to compare the predictions of our physics-based model with the results obtained using: (i) the distributed diffusion model \cite{Wu2016}, and (ii) a discrete simulation where the electrolyte is considered a separate domain. As shown in \cref{fig:Case1_DomainsA}, the metal domain is constrained at the bottom and subjected to an applied vertical displacement $U_{\mathrm{ext}}$ at the top edge, which results in the creation of a thin electrolyte layer of $h=U_{\mathrm{ext}}$. The explicit interface simulations directly simulate this domain using the method described in Ref. \cite{Hageman2022b}. For the phase field simulations, the electrolyte layer is replaced by the initial presence of the phase field variable, \cref{fig:Case1_DomainsB}, which is initialised using \cref{eq:initialization}. As the crack is stationary, the magnitude of $G_{c0}$ is taken to be sufficiently high ($2 \cdot 10^{10}$ J/m$^2$) to prevent the phase field from spreading past the region initialised through the history field. On the left side of the domain, constant concentrations and zero electrolyte potential are imposed. The metal has dimensions $H=50\;\mathrm{mm}$, $L=5\;\mathrm{mm}$, and is discretised using quadratic Lagrangian elements of size $0.2\;\mathrm{mm}\;\times0.2\;\mathrm{mm}$. The temporal discretisation is performed using a backward Euler method with an initial time increment of $30\;\mathrm{s}$, increasing by $5\%$ each time step to simulate a total duration of $200\;\mathrm{hours}$. Simulations are performed for the following magnitudes of the applied displacement (and thus the electrolyte height): $U_{\mathrm{ext}}=10^{-4}, \; 10^{-3},\; 10^{-2}\;\mathrm{and} \; 10^{-1} \;\mathrm{mm}$. In the phase field-base simulations (physics-based and distributed diffusion), results are obtained for the following choices of phase field length scale $\ell = 0.5,\; 0.75,\; 1,\;\mathrm{and}\;1.25\;\mathrm{mm}$.\\

\begin{figure}
    \centering
    \includegraphics[width=13cm]{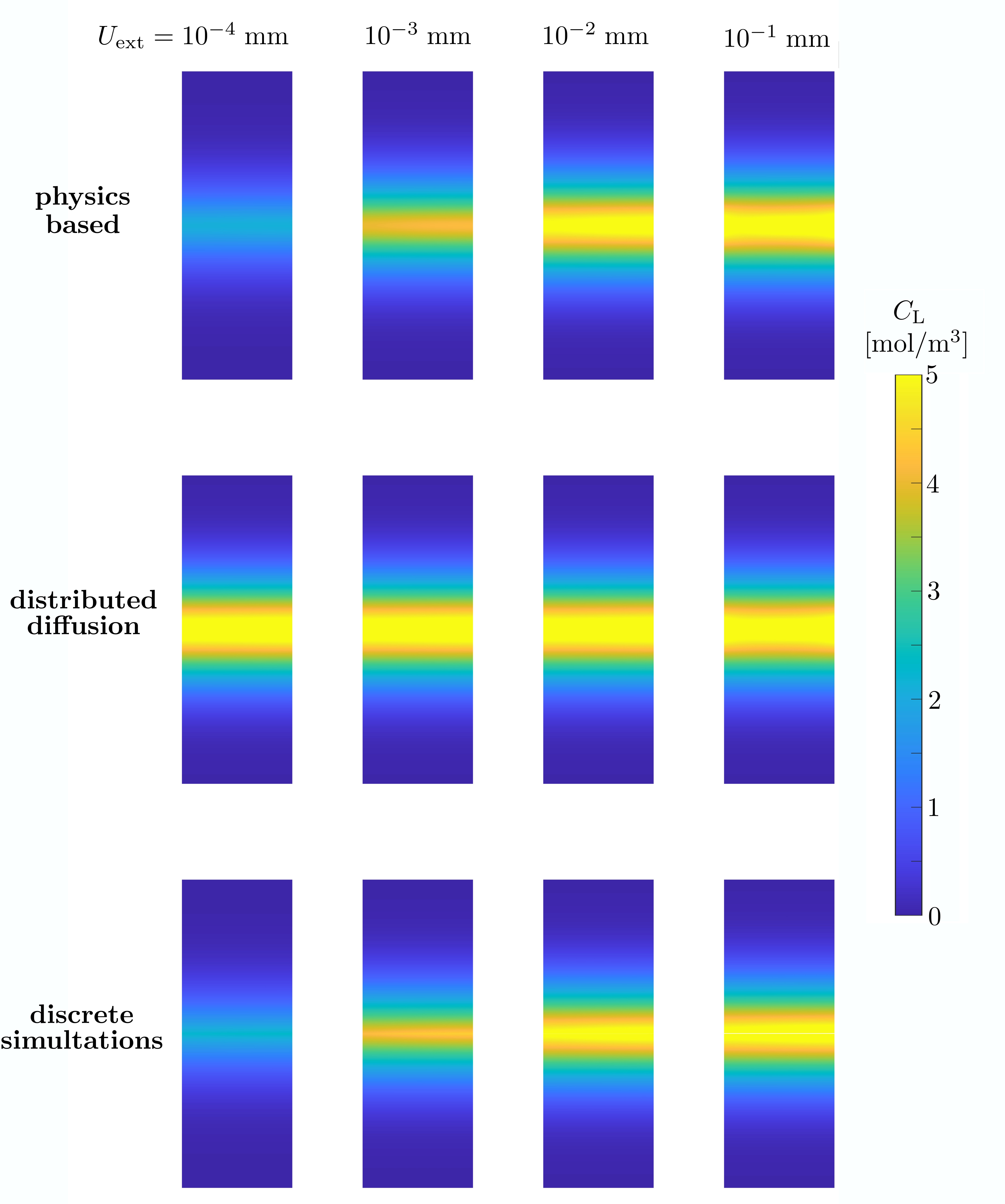}
    \caption{Lattice hydrogen concentration after $t=200\;\mathrm{hours}$ using $\ell=1\;\mathrm{mm}$, obtained from the physics-based, distributed diffusion, and discrete fracture simulations for varying $U_{\mathrm{ext}}$.}
    \label{fig:Case1_CL}
\end{figure}

\begin{figure}
    \centering
    \includegraphics[width=16cm]{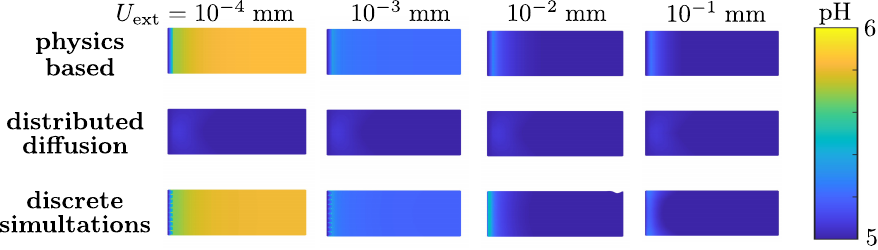}
    \caption{pH after $t=200\;\mathrm{hours}$ using $\ell=1\;\mathrm{mm}$, obtained from the physics-based, distributed diffusion, and discrete fracture simulations for varying $U_{\mathrm{ext}}$. pH only shown for locations where $\phi>0.01$.}
    \label{fig:Case1_pH}
\end{figure}

\begin{figure}
    \centering
    \includegraphics{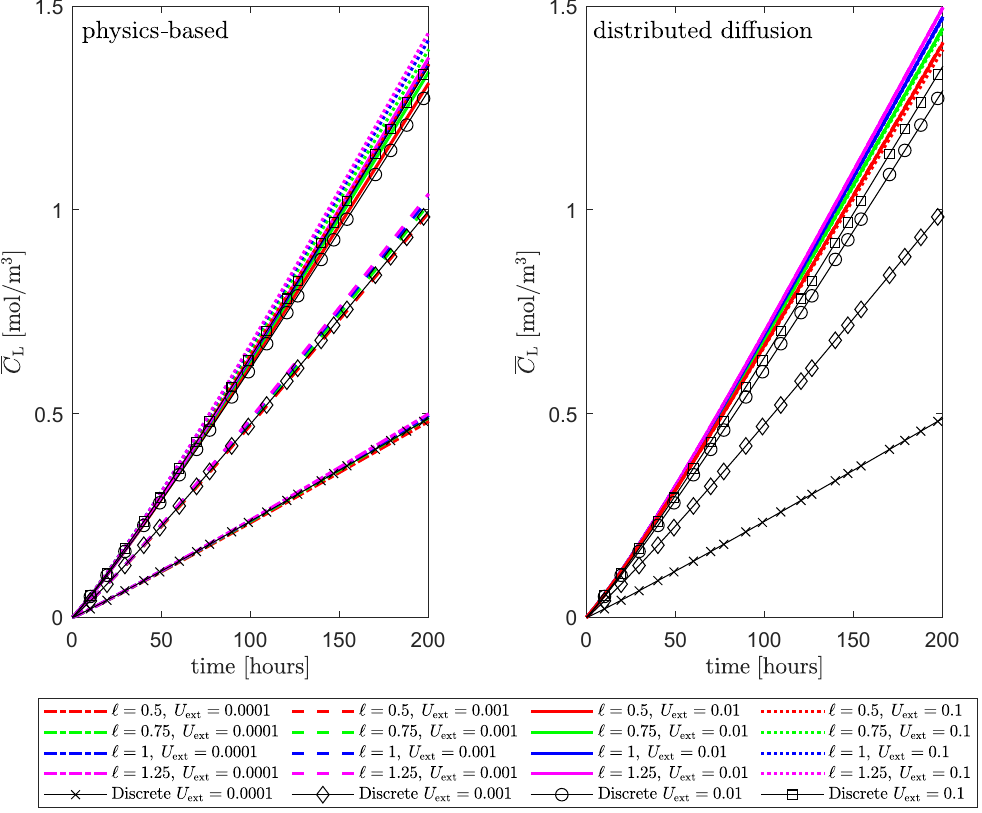}
    \caption{Evolution in time of the volume-averaged lattice hydrogen concentration for the physics-based (left) and distributed diffusion (right) models. Predictions are obtained for all combinations of applied displacement and phase field length scale considered, and these are compared with the discrete simulations (black lines with markers).}
    \label{fig:case1_CL_Avarage}
\end{figure}

The contours of interstitial lattice hydrogen concentration $C_\mathrm{L}$ calculated for the case of a phase field length scale $\ell=1\;\mathrm{mm}$ are shown in \cref{fig:Case1_CL}. It can be readily seen that the physics-based phase field formulation and the discrete simulations are in perfect agreement, showing a strong sensitivity to crack opening height (imposed through $U_{\mathrm{ext}}$). For small opening heights only a limited amount of hydrogen enters the metal, whereas for wider cracks significantly more hydrogen ingress takes place. The sensitivity to the crack geometry is more pronounced for small crack openings, with the hydrogen uptake predictions eventually saturating as $U_{\mathrm{ext}}$ increases, suggesting that there is an upper limit after which the opening height becomes less dominant in the hydrogen absorption process. This upper limit appears to correspond to the result obtained with the distributed diffusion model, which is unable to capture the smaller hydrogen uptake associated with smaller crack openings.\\

The electrolyte pH predictions, shown in \cref{fig:Case1_pH}, provide a similar qualitative picture. Here, the phase field-based predictions are again based on the choice of $\ell=1\;\mathrm{mm}$, and pH contours are given over a height equal to $U_{\mathrm{ext}}$ (the discrete electrolyte height, with figures being scaled for visibility purposes). The results are shown for a time of 200 hours. Again, the distributed diffusion model delivers crack height-insensitive results that appear to coincide with those associated with large crack openings. In contrast, the physics-based model obtains pH distributions similar to those of the discrete fracture simulations. Since the pH within the fracture directly influences the surface adsorbed hydrogen, an accurate estimation is paramount. Further results (not shown here) indicate that the agreement between the physics-based and discrete simulations also extends to the prediction of the concentration of other ionic species and the spatial distribution in electrolyte potential. In contrast, the distributed diffusion model is limited to characterising the environments intrinsic to high crack opening heights. One behaviour that the physics-based model is unable to capture is the two-dimensional distribution of the pH obtained for the $U_{\mathrm{ext}}=10^{-2}\;\mathrm{mm}$ simulations in the discrete and distributed diffusion models. These two models show a slight rise of the pH near the metal-electrolyte interface with a lower pH in the centre of the crack. In contrast, the physics-based model is built in such a way so as to enforce a zero concentration gradient in the direction normal to the crack. As a result, it is expected that the physics-based model starts to deviate from the direct simulation and distributed diffusion model for large opening heights, $h>>\mathcal{O}(1\;\mathrm{mm})$, where these two-dimensional effects dominate.\\ 

To quantify the behaviour of the system over time, we use the volume-averaged interstitial lattice hydrogen concentration, $\overline{C}_\mathrm{L}=\int C_\mathrm{L} \mathrm{d}\Omega / \int 1 \mathrm{d} \Omega$. The evolution of this average hydrogen concentration is shown in \cref{fig:case1_CL_Avarage} for all combinations of fracture model, applied displacement, and phase field length scale considered. For low imposed displacements, the physics-based model obtains a near perfect match with the discrete simulation result, independently of the phase field length scale adopted. As the crack opening height increases, the results start showing some sensitivity to the choice of phase field length scale, with smaller values of $\ell$ providing the most accurate results in terms of hydrogen uptake. However, even for the largest imposed displacement, the physics-based model reproduces the temporal behaviour correctly. In contrast, the distributed diffusion model overestimates the total hydrogen entry for all cases, with its results being independent of the imposed displacement and having a similar dependence on the length scale as the physics-based model. It can thus be concluded that the physics-based model presented here is a suitable strategy to endow phase field models with the ability of accurately predicting the electrolyte-crack interplay, capturing the sensitivity to the crack geometry.

\subsection{Electrolyte-driven crack propagation}
\label{sec:case2}

\begin{figure}
    \centering
    \includegraphics[width=5cm]{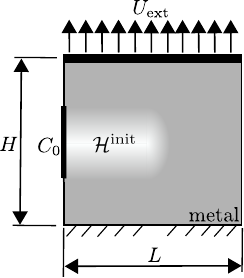}
    \caption{Electrolyte-driven crack propagation case study: geometry and boundary conditions.}
    \label{fig:Case2_Domain}
\end{figure}

The second case study aims at assessing the ability of the model in predicting the growth of cracks that contain aqueous electrolytes. To this end, we consider a square domain of dimensions $10\;\mathrm{mm}\;\times10\;\mathrm{mm}$ with an initial crack of length $5\;\mathrm{mm}$, as shown in \cref{fig:Case2_Domain}. This is a paradigmatic benchmark in the phase field fracture community \cite{Miehe2010b}. 
The square domain is discretised using a uniform mesh with  the element dimensions being $0.1\;\mathrm{mm}\times0.05\;\mathrm{mm}$. A constant external displacement $U_{\mathrm{ext}}=0.01\;\mathrm{mm}$ is imposed on the top edge, with this displacement being insufficient to cause the crack to propagate by itself. Over time, hydrogen is absorbed within the metal, reducing the material toughness and allowing the crack to propagate. The combination of imposed displacement and fracture energy has been selected such that no significant propagation occurs in the absence of hydrogen, while modest amounts of hydrogen ingress cause the domain to fully fracture. To track the evolution of these fractures, the total crack length is estimated based on the phase field distribution function, \cref{eq:distributor_phasefield}, such that:
\begin{equation}
    a = \int_{\Omega} \frac{\phi^2}{2\ell}+\frac{\ell}{2}|\bm{\nabla}\phi|^2\;\mathrm{d}\Omega
\end{equation}
While this does not provide the exact length over which the crack has propagated, it provides a good indication of the rate at which it evolves. Both the physics-based model and the distributed diffusion model are used to simulate the case, using phase field length scales $\ell = 0.125$, 0.25, 0.375 and 0.5 mm. Due to the difficulties of discretising the interior of a moving crack, no discrete fracture simulations were performed.\\ 

\begin{figure}
    \centering
    \begin{subfigure}[b]{\textwidth}
    \centering
    \includegraphics[clip, trim={0 50 0 0}]{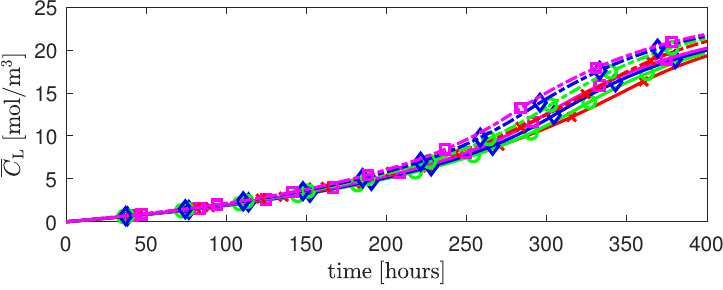}
    \caption{}
    \label{fig:case2_CL}
    \end{subfigure}
    \begin{subfigure}[b]{\textwidth}
    \centering
    \includegraphics{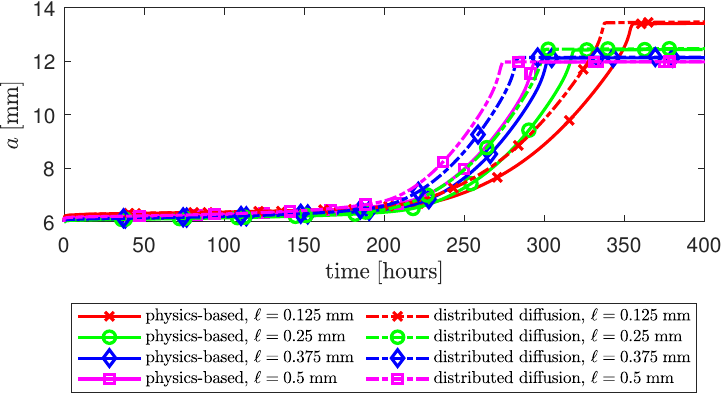}
    \caption{}
    \label{fig:case2_LFrac}
    \end{subfigure}
    \caption{Electrolyte-driven crack propagation case study. Predictions in time of: (a) the volume-averaged interstitial lattice hydrogen concentration uptake, and (b) the estimated fracture length. Results obtained for both the physics-based and the distributed diffusion models for handling electrolytes within cracks.}
    \label{fig:case2}
\end{figure}
\begin{figure}
     \centering
     \begin{subfigure}[b]{0.49\textwidth}
         \centering
         \includegraphics[width=8cm]{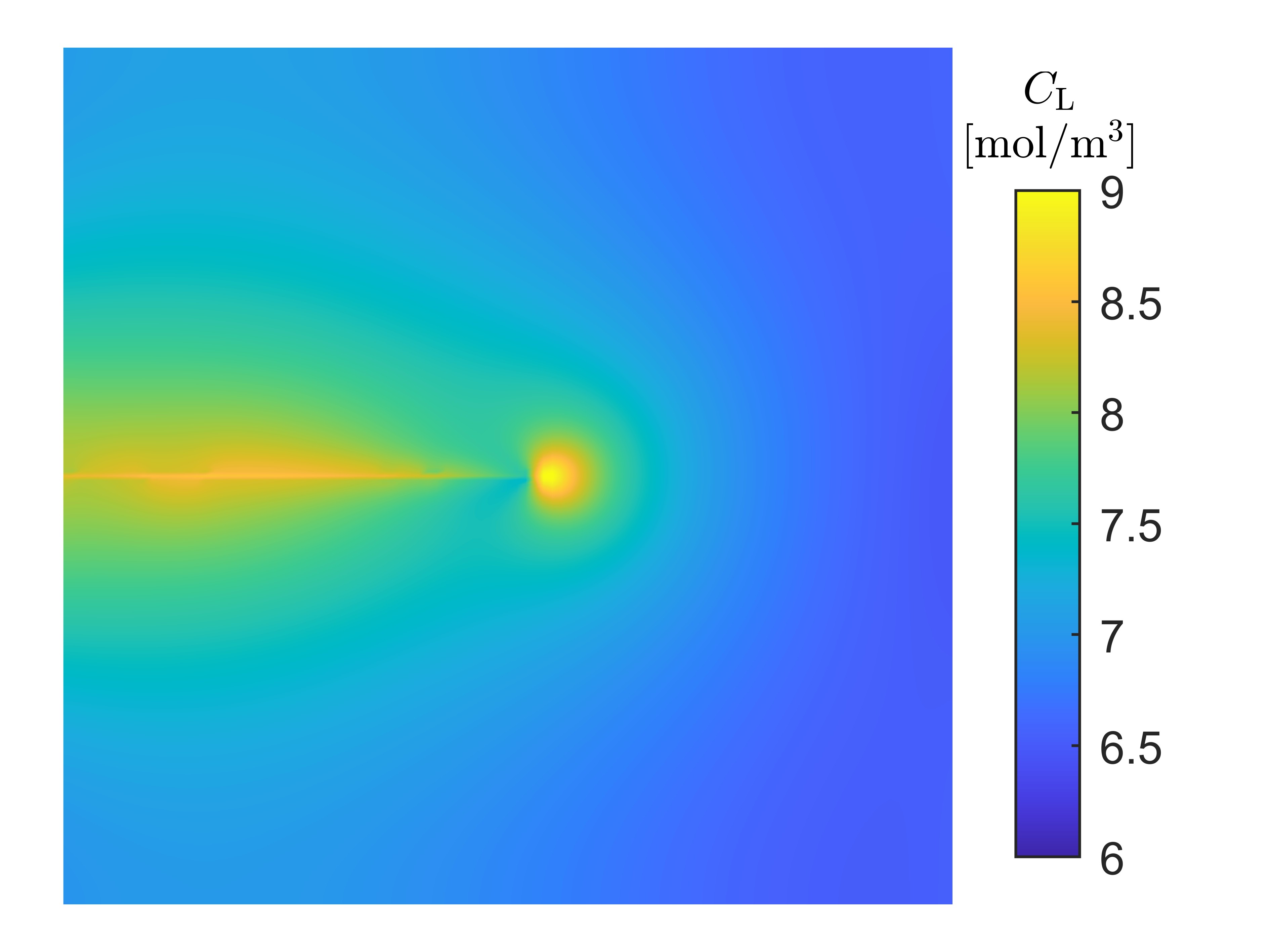}
         \caption{$\ell=0.125\;\mathrm{mm}$}
         \label{fig:Case2_CL_Surfs1}
     \end{subfigure}
     \begin{subfigure}[b]{0.49\textwidth}
         \centering
         \includegraphics[width=8cm]{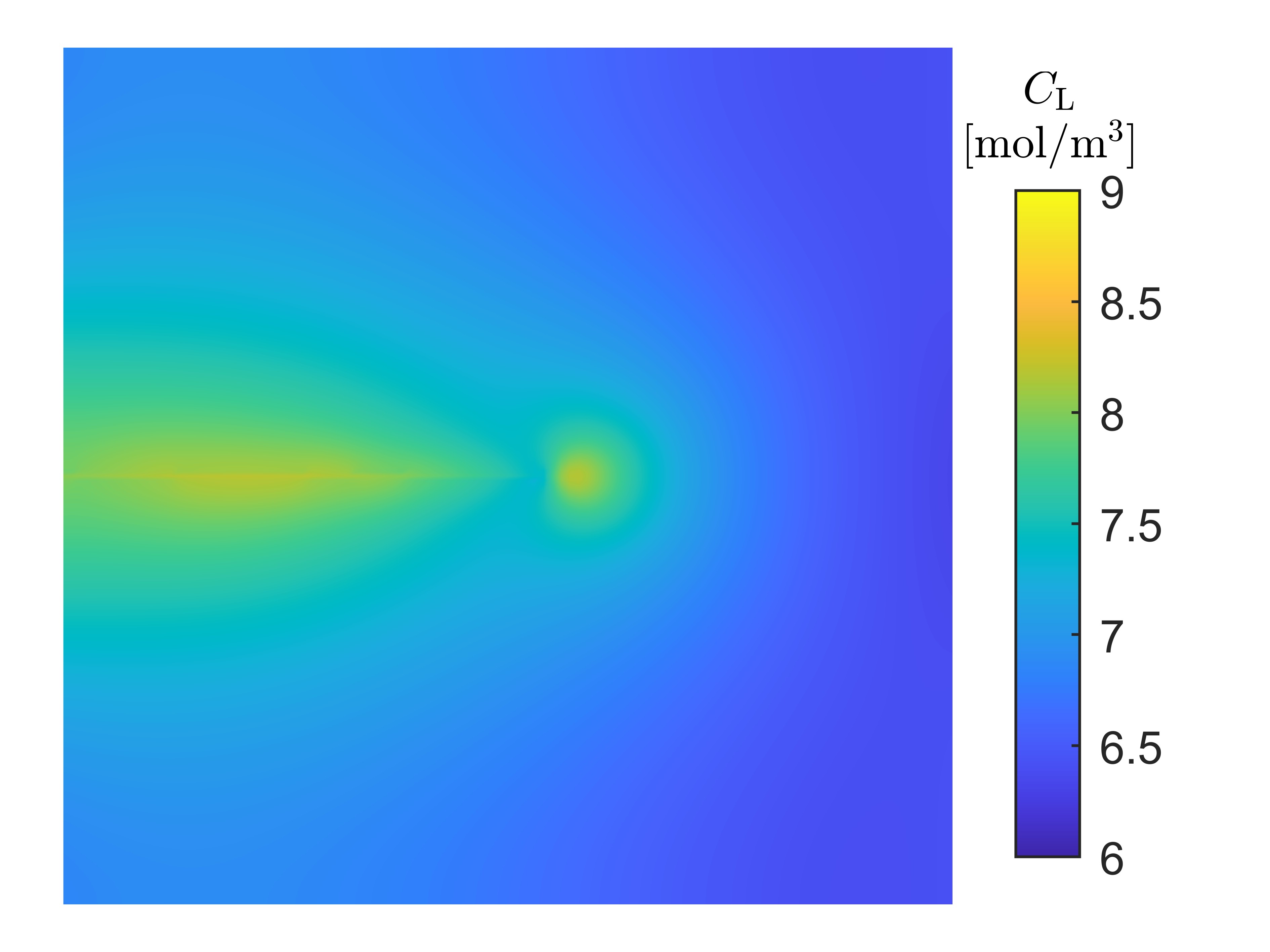}
         \caption{$\ell=0.25\;\mathrm{mm}$}
         \label{fig:Case2_CL_Surfs2}
     \end{subfigure}
    \caption{Electrolyte-driven crack propagation case study. Contours of lattice hydrogen concentration after $t=240\;\mathrm{hours}$. Results obtained with the physics-based model for two choices of phase field length scale: (a) $\ell=0.125$ mm, and (b) $\ell=0.25$ mm.}
    \label{fig:Case2_CL_Surfs}
\end{figure}

The results obtained are shown in \cref{fig:case2}, in terms of the evolution in time of the volume-averaged interstitial hydrogen concentration and of the crack length. As shown in Fig. \ref{fig:case2}a, and in agreement with expectations, the distributed diffusion model shows a larger hydrogen uptake initially, compared to the physics-based model. As a result of this higher uptake, the crack propagates sooner for the distributed diffusion model simulations, see \cref{fig:case2_LFrac}. Since the displacement on the top surface is constant throughout the simulation, this crack develops solely due to the role of hydrogen in reducing the fracture resistance of the material. In contrast to the static crack case from \cref{sec:case1}, this case shows a strong length-scale dependence for both the physics-based and distributed-diffusion results. The results show some sensitivity to the choice of phase field length scale, for both the distributed diffusion and physics-based models, with larger values of $\ell$ leading to earlier failures. This can be rationalised as follows. First, note that the choice of phase field length scale determines the strength of the material, as evident from the critical stress obtained for a one-dimensional solution of the phase-field problem, $\sigma_\mathrm{c} = 9/16 \sqrt{EG_\mathrm{c}/(6\ell)}$ \citep{Borden2012}. Although the boundary value problem under consideration involves a long crack (and thus toughness-dominated behaviour is expected \cite{Kristensen2021}), the magnitude of the strength imposes an upper limit on the hydrostatic stress levels that can be attained, and these govern hydrogen uptake. For example, under steady state conditions, the lattice hydrogen concentration reads,
\begin{equation}
    C_\mathrm{L} = C_\mathrm{0} \exp \left( \frac{\overline{V}_\mathrm{H}\sigma_\mathrm{H}}{RT} \right)
\end{equation}
\noindent where $C_\mathrm{0}$ is the reference, far-field hydrogen lattice concentration. The interplay between the material strength and the hydrogen localisation is shown in \cref{fig:Case2_CL_Surfs}, where contours of lattice hydrogen concentration are shown for two values of the phase field length scale, after a time of 240 hours. The results show how decreasing the magnitude of $\ell$ (i.e., increasing $\sigma_\mathrm{c}$) results in higher levels of interstitial hydrogen. Notably, this length scale dependence becomes more pronounced after the onset of crack growth. The damaged region is larger for higher $\ell$ values, providing a larger region of exposure to the hydrogen-containing electrolyte. This can be seen in Fig. \ref{fig:case2}a, where the differences between the predictions obtained with different $\ell$ values are seen to increase with time. These results confirm the ability of the proposed scheme to capture the influence of hydrogen uptake on propagating cracks.

\subsection{Coupling free-flowing and crack-contained electrolytes}
\label{sec:case3}

\begin{figure}
    \centering
    \includegraphics{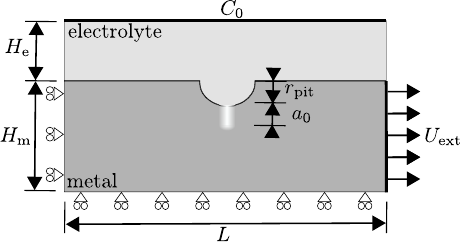}
    \caption{Coupling free-flowing and crack-contained electrolytes: Overview of the geometry and boundary conditions.}
    \label{fig:Domain_Case3}
\end{figure}
\begin{figure}
    \centering
    \begin{subfigure}[b]{0.49\textwidth}
        \centering
        \includegraphics[width=8cm]{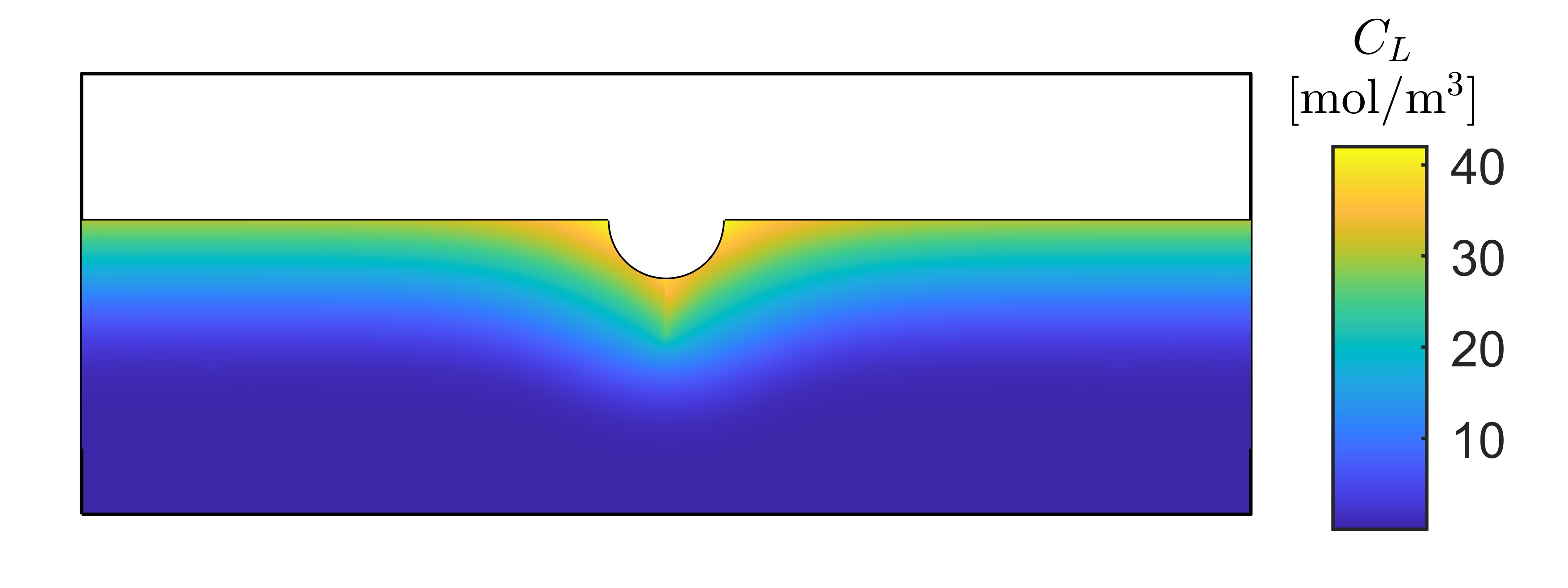}
        \caption{$E_\mathrm{m}=-0.4\;\mathrm{V}_{\mathrm{SHE}}$}
        \label{fig:case3_CL_-04}
    \end{subfigure}
    \begin{subfigure}[b]{0.49\textwidth}
        \centering
        \includegraphics[width=8cm]{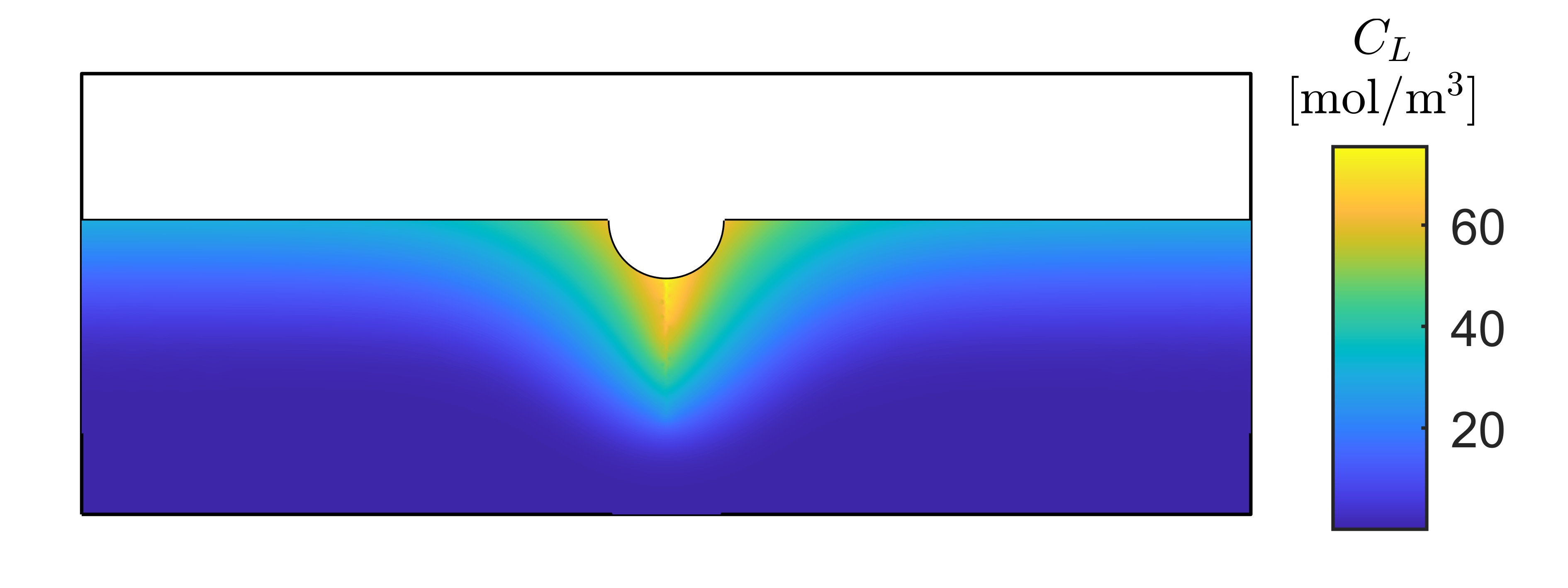}
        \caption{$E_\mathrm{m}=0.2\;\mathrm{V}_{\mathrm{SHE}}$}
        \label{fig:case3_CL_02}
    \end{subfigure}
    \caption{Coupling free-flowing and crack-contained electrolytes: contours of interstitial lattice hydrogen concentration obtained after $t=10\;\mathrm{hours}$. Results are presented for two choices of applied potential: (a) $E_\mathrm{m}=-0.4\;\mathrm{V}_{\mathrm{SHE}}$, and (b) $E_\mathrm{m}=0.2\;\mathrm{V}_{\mathrm{SHE}}$.}
    \label{fig:case3_CL}
\end{figure}
\begin{figure}
    \centering
    \begin{subfigure}[b]{0.49\textwidth}
        \centering
        \includegraphics[width=8cm]{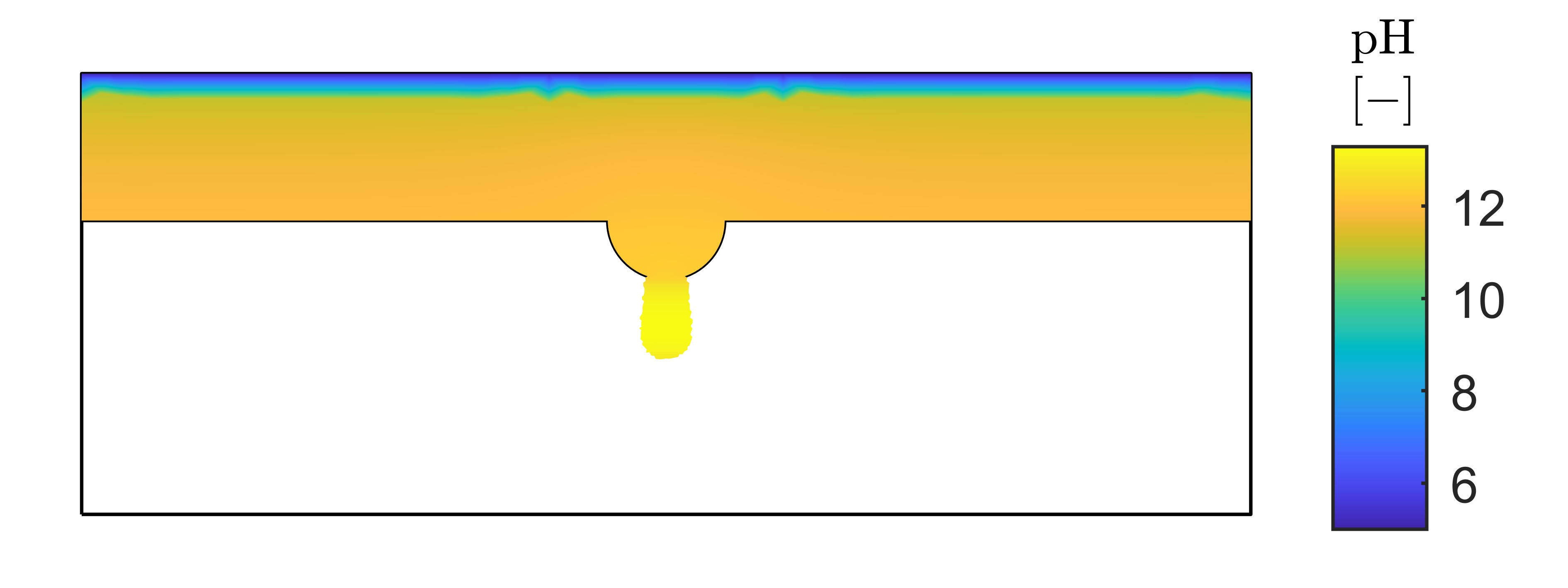}
        \caption{$E_\mathrm{m}=-0.4\;\mathrm{V}_{\mathrm{SHE}}$}
        \label{fig:case3_PH_-04}
    \end{subfigure}
    \begin{subfigure}[b]{0.49\textwidth}
        \centering
        \includegraphics[width=8cm]{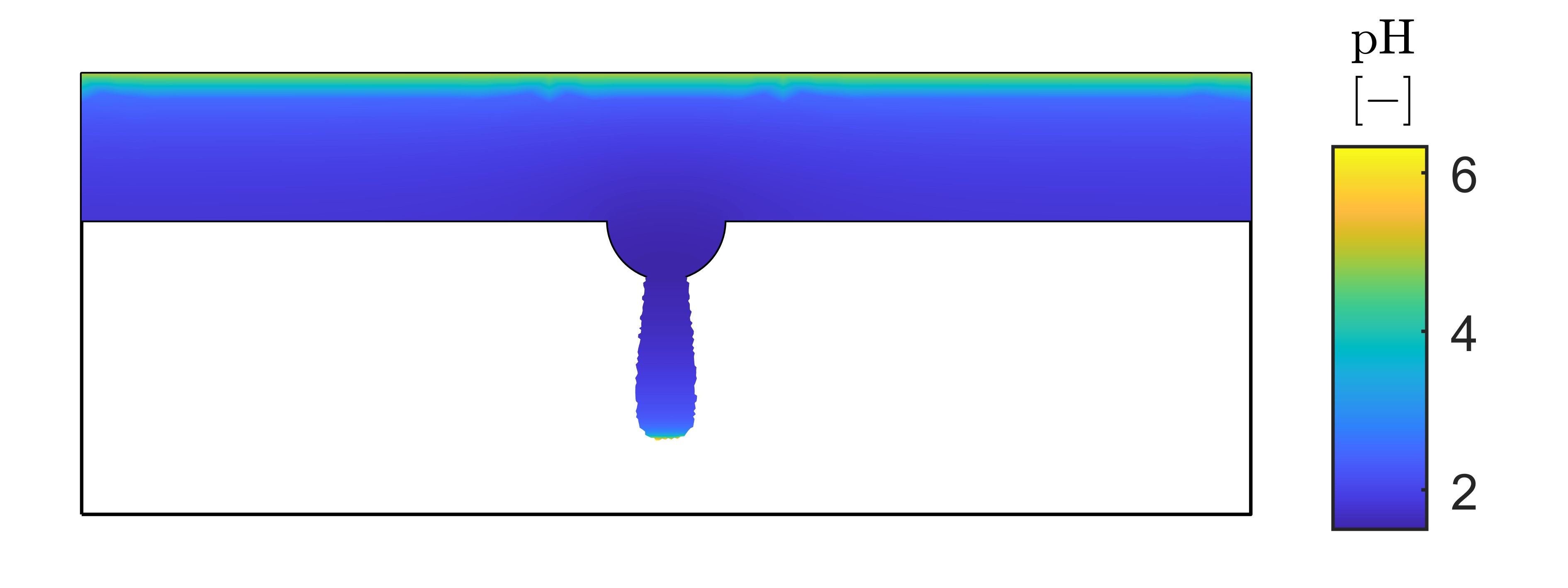}
        \caption{$E_\mathrm{m}=0.2\;\mathrm{V}_{\mathrm{SHE}}$}
        \label{fig:case3_PH_02}
    \end{subfigure}
    \caption{Coupling free-flowing and crack-contained electrolytes: pH contours obtained after $t=10\;\mathrm{hours}$. Results are presented in regions where $\phi>0.1$ and for two choices of applied potential: (a) $E_\mathrm{m}=-0.4\;\mathrm{V}_{\mathrm{SHE}}$, and (b) $E_\mathrm{m}=0.2\;\mathrm{V}_{\mathrm{SHE}}$.}
    \label{fig:case3_PH}
\end{figure}
\begin{figure}
    \centering
    \begin{subfigure}[b]{\textwidth}
        \centering
        \includegraphics[clip, trim={0 40 0 0}]{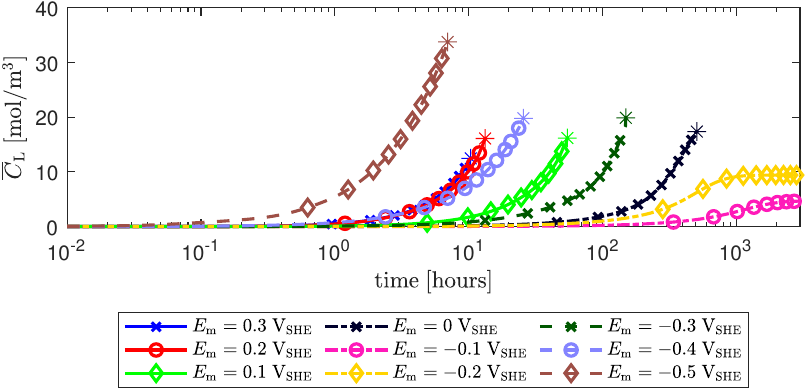}
        \caption{}
        \label{fig:case3a}
    \end{subfigure}
    \begin{subfigure}[b]{\textwidth}
        \centering
        \includegraphics{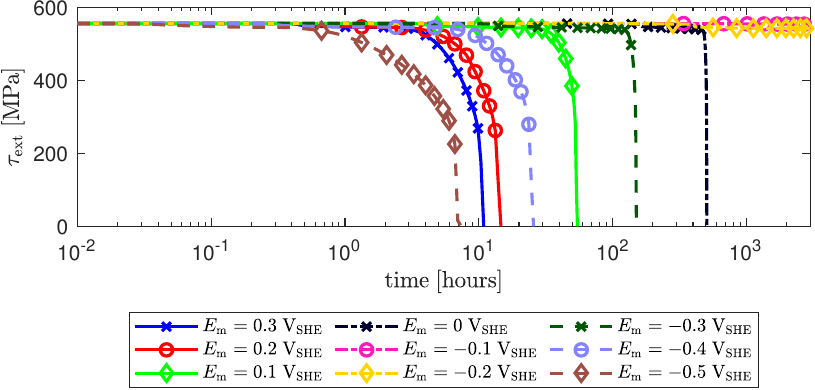}
        \caption{}
        \label{fig:case3b}
    \end{subfigure}
    \caption{Coupling free-flowing and crack-contained electrolytes. Predictions of the evolution in time of (a) the volume-averaged interstitial lattice hydrogen concentration $\overline{C}_\mathrm{L}$, and (b) the external traction, indicating the loss of load carrying capacity due to cracking. Results are obtained for a wide range of applied potentials $E_\mathrm{m}$. In Fig. \ref{fig:case3a}, the star markers denote the moment of failure.}
    \label{fig:case3}
\end{figure}

The last case study addresses the most general scenario, one where there is a separate, free-flowing electrolyte domain, in addition to a solid domain and an electrolyte-containing crack domain. A sketch of the boundary value problem under consideration is shown in \cref{fig:Domain_Case3}. The boundary value problem involves a metal of size $L\times H_\mathrm{m} = 4\times 1 \;\mathrm{cm}$, containing a pit in the centre with a radius of $r_{\mathrm{pit}}=2\;\mathrm{mm}$, and an initial crack of length $a_{\mathrm{0}}=1\;\mathrm{mm}$. On top of this metal, an electrolyte layer of height $H_\mathrm{e}=5\;\mathrm{mm}$ is present. This electrolyte is simulated directly through the Nernst-Planck equations; \cref{eq:NernstPlanck,eq:electroneutrality}. It should be noted that, while the fracture-contained electrolyte takes the displacements of the metal into account through the crack opening height, the free electrolyte does not include any effects resulting from geometry changes. On the right side of the metal, a constant displacement $U_\mathrm{ext}=0.02\;\mathrm{mm}$ is imposed. To demonstrate the ability of the model in capturing the sensitivity to the external environment, simulations are conducted for a wide range of applied metal potentials, going from $E_\mathrm{m}=-0.5\;\mathrm{V}_{\mathrm{SHE}}$ to $E_\mathrm{m}=0.3\;\mathrm{V}_{\mathrm{SHE}}$. Lower potentials, typical of cathodic protection conditions, strongly accelerate hydrogen reactions, while higher potentials enhance the corrosion rate. The domain is discretised using quadratic triangular elements, using small elements with characteristic size of $0.1\;\mathrm{mm}$ near the expected crack path, and larger elements up to $1\;\mathrm{mm}$ for the electrolyte and metal away from the crack. This results in a total of 37,000 nodes, with a total of 368,000 degrees of freedom. \\

Representative results for the spatial distribution of the interstitial hydrogen concentration are shown in \cref{fig:case3_CL}. The results are given for a time of $t=10$ hours and two choices of applied potential ($E_\mathrm{m}=-0.4\;\mathrm{V}_{\mathrm{SHE}}$ and $E_\mathrm{m}=0.2\;\mathrm{V}_{\mathrm{SHE}}$). While for negative potentials the hydrogen is absorbed from both the exterior and crack surfaces, the majority of the hydrogen for the positive potential simulation enters through the crack. These results can be rationalised by inspecting the pH contours obtained, which are shown in \cref{fig:case3_PH}. Here, it is worth noting that although the pH is calculated in the entire domain, its physical meaning is limited to electrolyte-containing regions and thus results are only shown for regions where $\phi>0.1$. The calculations show that the pH is dominated by the hydrogen evolution reactions for negative metal potentials (\cref{fig:case3_PH}a), causing the electrolyte to become highly basic within the defect. In contrast, the accelerated corrosion occurring for positive metal potentials lowers the pH within the pit and crack regions relative to the exterior surface. As a result of this low pH, hydrogen uptake is enhanced within the crack for the applied potential $E_\mathrm{m}=0.2\;\mathrm{V}_{\mathrm{SHE}}$, while a smaller sensitivity to the existence of defects is observed for $E_\mathrm{m}=-0.4\;\mathrm{V}_{\mathrm{SHE}}$.\\

The volume-averaged lattice hydrogen concentration $\overline{C}_\mathrm{L}$ and the associated loss of load carrying capacity are given in \cref{fig:case3}, as a function of time. Results are given for a wide range of applied potentials so as to showcase the ability of the model in predicting the sensitivity to the environment of the hydrogen uptake and the failure time. As shown in \cref{fig:case3b}, the initial external load is not sufficient to cause an immediate fracture of the metal and thus crack growth requires the accumulation of sufficient hydrogen in the crack tip region. For the $E_\mathrm{m}=-0.2\;\mathrm{V}_\mathrm{SHE}$ and $E_\mathrm{m}=-0.1\;\mathrm{V}_\mathrm{SHE}$ cases, no crack propagation occurs within $100\;\mathrm{days}$ and the average interstitial lattice hydrogen (\cref{fig:case3a}) starts to plateau, indicating that under these circumstances (environment, material, applied load) no fracture propagation due to hydrogen embrittlement will ever occur. For all other metal potentials, the hydrogen absorption is sufficient to cause a crack to develop, reducing the external traction to zero. Comparing Figs. \ref{fig:case3}a and \ref{fig:case3}b one can see the role of localised hydrogen uptake in driving the cracking process. The cases dominated by corrosion (low $E_\mathrm{m}$) show a lower average hydrogen concentration at failure, as for these cases the majority of the hydrogen is absorbed into the metal through the crack walls near the fracture tip. In contrast, the cases with a negative metal potential absorb the majority of hydrogen from the exterior surface, away from the crack tip and stress concentrations, and thus require this hydrogen to diffuse towards the crack. This results in a higher average hydrogen concentration at the point of failure. It can be seen that the most aggressive environment corresponds to the one with the lowest applied potential ($E_\mathrm{m}=-0.5\;\mathrm{V}_\mathrm{SHE}$), as the hydrogen reactions are greatly enhanced. However, the interplay between applied potential and hydrogen embrittlement susceptibility is not straightforward. As discussed in the context of Figs. \ref{fig:case3_CL} and \ref{fig:case3_PH}, the enhanced corrosion process associated with positive applied potentials can lead to a localised reduction in pH, and thus an increase in hydrogen uptake, which can overcompensate the reduction in hydrogen uptake associated with the deceleration in hydrogen reaction rates. The results obtained not only showcase the ability of the model to shed light into the complex interplay between the environment and electro-chemo-mechanical failures but also demonstrate its potential in delivering predictions over technologically-relevant scales, despite the large number of degrees-of-freedom involved (12 per node).

\section{Conclusions}
\label{sec:conclusion}
We have presented a new phase field-based theoretical and computational framework for simulating electro-chemo-mechanical fracture. For the first time, the modelling framework combines: (i) an electrochemical description of electrolyte behaviour, capable of handling an arbitrary number of ionic species and changes in electrolyte potential, (ii) surface reaction modelling at the electrolyte-electrode interface, (iii) species absorption and subsequent stress-driven bulk diffusion within the electrode metal, and (iv) a phase field description of fracture that incorporates toughness degradation due to the presence of aggressive species. Moreover, we present a novel formulation to represent the electrolyte contained within cracks within the context of phase field fracture models. This formulation is based upon the governing equations for the electrolyte, mapping from an electrolyte represented in a discrete manner to a smeared representation of the electrolyte. This approach is compared to the the widely used distributed diffusion model, showing that both can be described through similar schemes, only altering the capacity, surface, and diffusion distribution functions. The theoretical framework was implemented using the finite element method. The coupled electrical-chemical-deformation-fracture problem was solved in a staggered manner, with the primary fields (nodal degrees-of-freedom) being: (i) the electrolyte potential, (ii) the concentrations of relevant ionic species, (iii) the interface coverage of absorbable species, (iv) the concentration of diluted species in the bulk metal, (v) the displacement field, and (vi) the phase field order parameter. Given the number of fields involved, special emphasis is placed on improving stability and efficiency. Among others, we introduce a lumped integration scheme that greatly reduces oscillations and enables adopting large time increments without convergence problems. Also, strategies are adopted to prevent ill-constrained degrees-of-freedom. To demonstrate the potential of our computational framework we particularise our generalised model to the analysis of metallic fracture due to the uptake of hydrogen from aqueous electrolytes, a technologically-relevant problem that is pervasive across the defence, transport, construction and energy sectors. Several boundary value problems are addressed to showcase the ability of the model to adequately simulate the behaviour of electrolytes contained within cracks and to capture the interplay between fracture and electro-chemo-mechanical phenomena. Key findings include:
\begin{itemize}
    \item The physics-based formulation presented to describe electrolytes within cracks is shown to capture the sensitivity to crack opening height, unlike other existing models. Predictions of hydrogen uptake and ionic species distribution show an excellent agreement with discrete fracture simulations. Moreover, the predictions of this physics-based model display a negligible sensitivity to the choice of phase field length scale $\ell$ for stationary cracks, with some sensitivity being observed in propagating cracks due to the relation between $\ell$ and the material strength.
    \item The model is shown to be capable of adequately predicting the interplay between the environment, the material properties and the applied load for both crack-contained electrolytes and the more general case of free-flowing electrolytes. Widely observed experimental trends can now be rationalised in terms of changes in electrolyte behaviour, hydrogen uptake and toughness degradation. 
    \item The analysis of defect-containing metals exposed to free-flowing, hydrogen-containing electrolytes reveals that high applied potentials, which favour corrosive reactions relative to the hydrogen evolution reaction, can result in early failures due to local acidification of the electrolyte solution in the defect region. 
\end{itemize}

\section*{Acknowledgments}
\noindent Financial support through grant EP/V009680/1 (``NEXTGEM") from the UK Engineering and Physical Sciences Research Council (EPSRC) is gratefully acknowledged. Tim Hageman additionally acknowledges support through the research fellowship scheme of the Royal Commission for the Exhibition of 1851, and Emilio Mart\'{\i}nez-Pa\~neda additionally acknowledges financial support from UKRI's Future Leaders Fellowship programme [grant MR/V024124/1]. The authors also acknowledge computational resources and support provided by the Imperial College Research Computing Service (http://doi.org/10.14469/hpc/2232).

\section*{Data availability}
\noindent The \texttt{MATLAB} code used to produce the results presented in this paper, together with documentation detailing the use of this code, are made freely available at \url{www.imperial.ac.uk/mechanics-materials/codes} and \url{www.empaneda.com}. Documentation is also provided, along with example files that enable reproduction of the results shown in  \cref{sec:case1,sec:case2}.

\appendix

\end{document}